\documentclass[twocolumn,twoside,a4paper,10pt]{IEEEtran}
\makeatletter
\def\ps@headings{%
\def\@oddhead{\mbox{}\scriptsize\rightmark \hfil \thepage}%
\def\@evenhead{\scriptsize\thepage \hfil \leftmark\mbox{}}%
\def\@oddfoot{}%
\def\@evenfoot{}}
\makeatother
\usepackage{graphicx, graphics,epsfig,subfigure}
\usepackage{float}
\usepackage{color, cite}
\usepackage{amssymb, amsmath}
\usepackage{amsthm}
\usepackage{algorithm}
\usepackage{algorithmic}
\usepackage{amsfonts, dsfont}
\usepackage{comment}
\usepackage{cases}
\usepackage{tabularx}
\usepackage{latexsym}
\usepackage{colortbl}

\usepackage{txfonts}
\ifodd 1
\newcommand{\com}[1]{\textbf{\color{blue} (COMMENT: #1)}} 
\else

\newcommand{\com}[1]{}
\fi

\newcommand{\beq}   {\begin{equation}}
\newcommand{\eeq}   {\end{equation}}
\newcommand{\bea}   {\begin{eqnarray}}
\newcommand{\eea}   {\end{eqnarray}}
\newcommand{\bda}   {\begin{eqnarray*}}
\newcommand{\eda}   {\end{eqnarray*}}
\newcommand{\bdalign}   {\begin{align*}}
\newcommand{\edalign}   {\end{align*}}

\newcommand {\btau}{\mbox{\boldmath $\tau$}}

\newcommand {\br}{\mbox{\boldmath $r$}}

\newcommand {\bx}{\mbox{\boldmath $x$}}
\newcommand {\by}{\mbox{\boldmath $y$}}

\newtheorem{theorem}{{\bf Theorem}}

\title{\Huge \bf TCP Reno over Adaptive CSMA}
\author{
Wei Chen$^*$, Yue Wang$^{\dagger}$, Minghua Chen$^*$, and Soung Chang Liew$^*$ \\
$^*$Dept. of Information Engineering, The Chinese University of Hong Kong,\\
$^{\dagger}$School of Information, Central University of Finance and Economics \\
$^*$\{cw008, minghua, soung\}@ie.cuhk.edu.hk, $^{\dagger}$yue.wang@cufe.edu.cn
}

\begin{document}

\maketitle

\begin{abstract}
An interesting distributed adaptive CSMA MAC protocol, called adaptive CSMA, was proposed
recently to schedule any strictly feasible achievable rates inside the capacity region.
Of particular interest is the fact that the adaptive CSMA
can achieve a system utility arbitrarily close to that is achievable under
a central scheduler. However, a specially designed transport-layer rate
controller is needed for this result. An outstanding question is whether
the widely-installed TCP Reno is compatible with adaptive CSMA and can achieve the same result. The
answer to this question will determine how close to practical deployment
adaptive CSMA is. Our answer is yes and no. First, we observe that
running TCP Reno directly over adaptive CSMA results in severe starvation
problems. Effectively, its performance is no better than that of TCP
Reno over legacy CSMA (IEEE 802.11), and the potentials of adaptive
CSMA cannot be realized.
Fortunately, we find that multi-connection TCP Reno over adaptive CSMA
with active queue management can materialize the advantages of adaptive
CSMA. NS-2 simulations demonstrate that our solution can alleviate starvation
and achieve fair and efficient rate allocation. Multi-connection
TCP can be implemented at either application or transport layer. Application-layer
implementation requires no kernel modification, making the solution
readily deployable in networks running adaptive CSMA.
\end{abstract}


\section{Introduction}

\label{sec:intro} The Carrier Sense Multiple Access (CSMA) protocol
is widely used in local-area networks, including Wi-Fi. As the deployment
of Wi-Fi spreads, it is now common to find multiple co-located Wi-Fi
networks with partially overlapping coverage. In such networks,
each link can only sense a subset of other
links, and characterizing the link throughput achievable by CSMA is challenging.

Refs. \cite{wang2005throughput} and~\cite{liew1854back} show that
the saturated link throughput achieved by CSMA can be studied via a
time-reversible Markov chain and can be expressed in a product form.
Further, authors in \cite{liew1854back} showed that the product
form results are insensitive to the packet length and back-off time
distributions given the ratio of their means.

Based on these results on the CSMA link throughput, Jiang
and Walrand \cite{jiang2008distributed}, and later others \cite{liu2010towards,ni2010q,rajagopalan2008distributed,chen2010markov},
present a distributed adaptive CSMA (A-CSMA) scheme that can achieve
desired link throughput according to some system utility objective.
It is shown that the distributed A-CSMA scheme can achieve almost
the entire capacity region that can be achieved under a central scheduler.
Given its attractiveness and impenetrability~\cite{lee2009implementing},
an outstanding question is how close A-CSMA is to practical deployment.
As originally studied in~\cite{jiang2008distributed}, specially
designed transport-layer rate controllers were required to inter-work
with A-CSMA.

In view of the large installed base of TCP Reno, a question is whether
TCP Reno over A-CSMA works well.
We answer this question in this paper and make the following contributions:
\begin{itemize}
\item We observe that running TCP Reno directly over A-CSMA results in severe starvation problems.
Effectively, its performance is no better than that of TCP Reno over
legacy CSMA (IEEE 802.11) \cite{wang2005throughput}
\cite{liew1854back} \cite{rangwala2009neighborhood} \cite{tan2007congestion}.
Our analysis indicates that the root cause is that TCP Reno over A-CSMA
results in system instability.

\item We propose a multi-connection TCP Reno solution to inter-work with A-CSMA
in a compatible manner. Although multi-connection TCP has been explored before in the context of video transmission over cellular and wired networks \cite{chen2006flow,tullimas2008multimedia}, this paper applies it to solve starvation problem in TCP Reno over A-CSMA. Our solution is provably optimal under the
Network Utility Maximization (NUM) framework. Specifically, by adjusting
the number of TCP Reno connections for each session, any system utility
can be realized. NS-2 simulations demonstrate that this solution
can address starvation problems and achieve fair and efficient rate
allocation. The scheme is applicable to single-hop wireless networks, wired networks
as well as multihop networks with wired and wireless links.

\item Our multi-connection TCP Reno scheme can be implemented at either the transport
layer or the application layer. Application-layer implementation requires
no kernel modification to TCP Reno, and the solution can be readily
deployed over networks with A-CSMA.
\end{itemize}

The rest of the paper is organized as follows. We present related
work in Section~\ref{sec:related}, and the network model in Section~\ref{sec:model}.
We illustrate the starvation of TCP Reno A-CSMA in Section~\ref{sec:tcp.acsma}. We then propose our solution in the Section~\ref{sec:solu},
and evaluate its performance through NS-2 simulations in Section~\ref{sec:simul}.
Finally, Section~\ref{sec:conclu} concludes our paper.

\section{Related Work}

\label{sec:related}

Rate control (i.e., congestion control) is important for data transmission
over both wired and wireless networks. The widely deployed rate control
scheme is TCP Reno. To put our study in context, we classify the possible
solutions into four classes, as shown in Table~\ref{tab:classes},
where we refer to CSMA/CA scheme used in today's Wi-Fi
as legacy CSMA (L-CSMA).

Class $1$ solutions use TCP Reno over L-CSMA. There have been extensive
work studying the fairness problem of this scheme~\cite{wang2005throughput}~\cite{liew1854back}.
They reveal that TCP Reno over L-CSMA results in severe starvation
problem. The main reason is that L-CSMA is not an efficient scheduling
algorithm and it can only schedule a fraction of the capacity
region. 
%

\begin{table}[H]
 \caption{Rate control and Scheduling Solution Classes}
\label{tab:classes} \centering \begin{tabular}{|l|l|l|}
\hline
 & L-CSMA  & modified-MAC \tabularnewline
\hline
TCP Reno  & class 1  & class 2 \tabularnewline
\hline
modified TCP  & class 3  & class 4 \tabularnewline
\hline
\end{tabular}
\end{table}

Class $3$ solutions, for instance \cite{rangwala2009neighborhood}
\cite{tan2007congestion}, do not make any changes to the widely used
IEEE802.11 MAC. In contrast, they limit the design space to within the scope
of transport layer. In~\cite{rangwala2009neighborhood}, the authors
propose an AIMD-based rate control protocol called WCP Wireless Control
Protocol (WCP) to address the efficient and fair rate allocation over
L-CSMA networks. WCP halves the rates of other flows if they are within
interference range of the congested link. Hence, queues of other fast
transmitting links are driven to empty, giving the congested link
opportunity to transmit. However, WCP requires message passing among
the interfered links. Meanwhile, each link must monitor the Round-trip
time (RTT) of every flow that traverses it. EWCCP~\cite{tan2007congestion}
is designed to be proportional-fair by periodically broadcasting the
average queue information to coordinate among interference links.
These periodical broadcasted message consumes the precious wireless
bandwidth.

There have been many efforts belonging to class $4$. Cross-layer designs based on new TCP and new MAC are adopted ~\cite{jiang2008distributed,ni2010q,rajagopalan2008distributed,liu2010towards,lin2004joint,chen2005joint}.
Cross-layer designs can improve the overall system performance and
make interaction between layers more transparent, but raise the hurdle
to practical deployment.

Our solution belongs to class $2$. First, it is a layered approach
rather than a cross-layer one, reducing the implementation difficulty.
Second, since TCP Reno is widely installed in practice, to make a
network function well, it is preferable to avoid modifying TCP Reno.

Our solution is based on multiple connection TCP Reno. There has been
work on using multiple connections to transfer data. For example,
multiple TFRC connections for streaming video~\cite{chen2006multiple}
and multiple TCP Reno connections for data and multimedia streaming~\cite{chen2006flow,tullimas2008multimedia}
have been discussed. These works focus on wired networks and wireless
cellular networks. In contrast, our work focuses on CSMA networks
and therefore solves a different set of challenges, including the
spatial wireless inference that was not considered in~\cite{chen2006multiple,chen2006flow,tullimas2008multimedia}.

Our solution utilizes active queue management (AQM) in the MAC. AQM
techniques have been proposed to both alleviate congestion
control problems~\cite{floyd1993random} as well as to implement the
provisioning of quality of service~\cite{clark1998explicit}. The
classical example of an AQM policy is RED~\cite{floyd1993random}.
In our study, however, we apply a different AQM policy: the packet
dropping probability is proportional to queue length. In our work,
by applying this AQM, our scheme converges to the optimal solution.

\section{Problem Settings}

\label{sec:model}
In this section, we provide problem settings and necessary background on A-CSMA and TCP Reno.
We model a CSMA wireless network by a graph $G=(V,E)$, where $V$ is the set of nodes (i.e.,
wireless transmitters and receivers) and $E$ is the set of links.
We consider the scenario where multiple users communicate over the
wireless network. Each user is associated with a flow over a pre-defined
route between a source and a destination. The key notations used
in this paper are listed in Table \ref{tab:key.note}. We use bold
symbols to denote vectors and matrices of these quantities, e.g.,
$\bx=[x_{s},s\in S]$.

\begin{table}[hbt]
\caption{Key Notation}
\label{tab:key.note} \centering \begin{tabular}{l|l}
\hline
\textbf{Notation}  & \textbf{Definition} \tabularnewline
\hline
\hline
$V$  & the set of nodes \tabularnewline
$E$  & the set of links \tabularnewline
$\mathcal{S}$  & the set of source-destination pairs \tabularnewline
$x_{s}$  & rate of flow $s$ \tabularnewline
$T_{s}$  & round trip time of flow $s$ \tabularnewline
$n_{s}$  & number of connection on flow $s$ \tabularnewline
$y_{l}$  & average rate of link $l$ \tabularnewline
$\tau_{i}$  & time portion of independent set $i$ \tabularnewline
$r_{l}$  & transmission aggressiveness (TA) of link $l$\tabularnewline
$p_{l}$  & price of link $l$ \tabularnewline
$l\in i$  & link $l$ belongs to independent set $i$ \tabularnewline
$l\in s$  & flow $s$ passes through link $l$ \tabularnewline
$\mathcal{I}$  & the set of independent sets \tabularnewline
\hline
\end{tabular}
\end{table}

In CSMA wireless networks, transmitter applies a carrier-sensing
mechanism to prevent collisions. In particular, a transmitter
listens for an RF carrier before sending its data. If a carrier is
sensed, the transmitter defers its transmission attempt, enters a
countdown process in which it waits for a random amount of time before
making another transmission attempt. As such, two links are allowed
to transmit simultaneously only if the simultaneous transmissions are under
the CSMA carrier-sensing operations.

If not properly designed, simultaneous transmissions allowed by the
carrier-sensing operations may still interfere with each
other, resulting in the hidden-node problem~\cite{jiang2008improving}.
Hidden-node-free CSMA network designs are attractive, because, importantly,
their throughput analysis is tractable~\cite{wang2005throughput}~\cite{liew1854back}.
In particular, the stationary distribution of the system state is in a product form
that is fundamental to the design of adaptive CSMA \cite{jiang2008distributed}.
As studied by~\cite{chau2009capacity}, any CSMA wireless network can always be made to be hidden-node-free by setting the carrier sensing range properly.

In this paper, we study the problem of end-to-end rate control and
link scheduling in hidden-node-free CSMA wireless networks.

\subsection{Capacity Region of CSMA Wireless Networks}

For CSMA wireless networks, we use the conflict
graph approach in~\cite{jain2005impact} to model the relationship
of wireless link interference. We denote the conflict graph of $G$
by $G_{c}=(V_{c},E_{c})$. The vertices of $G_{c}$ correspond to
links $E$ in the communication graph $G$, i.e., $V_{c}=E$. An edge
between two vertices means that the two corresponding
links can carrier-sensing each other. An independent set corresponds to
a feasible schedule which is a set of links in the communication graph
that can be active simultaneously.

Let $\mathcal{I}$ denote the set
of all independent sets, $\tau_{i}$ denote the fraction of time the system is in schedule
$i\in\mathcal{I}$. Assuming that all links have unit capacity,
the capacity region of link rate is given by
\begin{equation}
\Pi=\{\by\geq0|\forall l,\; y_{l}\leq\sum_{i:l\in i}\tau_{i},\sum_{i\in\mathcal{I}}\tau_{i}=1,\btau\geq0\}.\label{eq:cap.reg}
\end{equation}
Note that finding all the independent sets is NP-hard
~\cite{baker1994npc}. Consequently, characterizing the capacity region in
(\ref{eq:cap.reg}) is thus very challenging, leaving alone attaining it in practice.

\subsection{Throughput-optimality of A-CSMA}

\label{sec:acsma}
Refs.~\cite{boorstyn1987throughput},~\cite{wang2005throughput}
and~\cite{liew1854back} show that the link scheduling in CSMA
networks can be studied via a Markov chain, in which the states are
all independent sets. In this model, the transmitter of link $l$ counts down an exponentially distributed random period of time with mean $\exp(-\beta r_l)$ before transmission. When the link $l$ starts a transmission, it keeps the channel for an exponentially distributed period of time with mean one\footnote{Without loss of generality, we normalize the mean of transmission time to one.}. This CSMA Markov chain converges to its product form stationary distribution, which is given by \begin{equation}p(i,\br)=\frac{\exp\left(\sum_{l\in i}\beta r_l\right)}{\sum_{i^{\prime}\in\mathcal{I}}\exp\left(\sum_{l\in i^{\prime}}\beta r_l\right)},\;\forall i\in \mathcal{I},\end{equation}
where $\beta$ is the a positive constant.
The average portion of time link $l$ is active, denoted by $y_{l}$, achieved in steady state is given by
\begin{equation}\label{eq:csma.stable.rate}y_l(\br)=\sum_{i:l\in i}p(i,\br)=\frac{\sum_{i:l\in i}\exp\left(\sum_{l\in i}\beta r_l\right)}{\sum_{i^{\prime}\in\mathcal{I}}\exp\left(\sum_{l\in i^{\prime}}\beta r_l\right)},\;\;\forall l\in E.\end{equation}

The authors in~\cite{jiang2008distributed} introduce a fully distributed adaptive CSMA (A-CSMA) algorithm as follows:
\newtheorem*{alg}{{\bf Algorithm A-CSMA}}
\begin{alg}
\begin{equation}\dot{r}_{l}=\alpha[a_{l}^{\prime}-d_{l}^{\prime}]_{r_{l}}^{+},\label{eq:dyn.r}\end{equation}
where $a_{l}^{\prime}$ and $d_{l}^{\prime}$ are the empirical average
arrival rate and service rate of link $l$, $\alpha$ is constant step
size and notationally,
\begin{equation}
[g(z)]_{z}^{+}=\left\{ \begin{array}{ll}
\max(g(z),0), & \mbox{if \ensuremath{z\leq0}}\\
g(z), & \mbox{otherwise}\end{array}\right.\end{equation}
\end{alg}

\begin{theorem}[\cite{jiang2008distributed}]
\label{the:acsma}
Assume arrival rate vector $\by^{\prime}$ is in interior of capacity region $\Pi$. Then with Algorithm A-CSMA,
$\br$ converges to $\br^*$, with probability one, where $\br^*$ satisfies that $y^{\prime}_{l}\leq y_l(\br^*)$.
\end{theorem}

\emph{Remark 1:} it is not difficult to verify that $r_{l}$ is proportional to its local queue length of link $l$.
Therefore, A-CSMA scheduling Algorithm A-CSMA can be implemented in a fully distributed way.
In A-CSMA, before transmitting, any link $l$ sets up a random exponentially distributed counter with mean equal to $\exp(-\beta r_l)$. When link $l$ obtains the channel, it transmits a packet with length which is exponentially distributed with mean equal to one. Note A-CSMA differs from L-CSMA in that the link's countdown timer
is a function of its queue length, as a result, A-CSMA gives priority to links with heavy loads. We remark that A-CSMA is simple to implement by slight modification to today's CSMA, and there have been efforts in doing so \cite{lee2009implementing}.

\emph{Remark 2:} Theorem \ref{the:acsma} shows that the product-form distribution $\by(\br)$ achieved by A-CSMA is larger than any interior feasible rate vector. This means A-CSMA stabilizes the system for all feasible arrival rate vector, i.e., A-CSMA scheduling algorithm is throughput-optimal. Therefore, A-CSMA has superior performance as compared to L-CSMA.

For more details of A-CSMA, interested readers can refer to the work in~\cite{jiang2008distributed,liu2010towards,ni2010q,rajagopalan2008distributed,chen2010markov}.

\subsection{TCP Reno Congestion Control Modeling\label{sec:TCP-Reno-Congestion}}

\label{sec:tcp}TCP Reno adjusts the flow rate by controlling the congestion window size $W$.
During the congestion avoidance phase, TCP Reno increments its congestion window
$W$ by one every round trip time if no loss is observed. It halves its window whenever packet loss is detected.

In~\cite{network.optimization.control}, the authors model this dynamic of the congestion
avoidance phase of TCP Reno with the following dynamic equation: \begin{equation}
\dot{x_{s}}=\frac{x_{s}^{2}}{2}\left(\frac{2}{T_{s}^{2}x_{s}^{2}}-\sum_{l\in s}p_{l}\right)\label{eq:tcp.rate}\end{equation}
where $T_{s}$ is round-trip-time(RTT) of flow $s$, and $p_{l}$
denotes the price (probability a packet will be dropped) of link $l$.
When link $l$ applies drop-tail queue, \begin{equation}
p_{l}=\frac{(\sum_{s:l\in s}x_{s}-y_{l})^{+}}{\sum_{s:l\in s}x_{s}},\label{eq:link.price}\end{equation}
 where $y_{l}$ denotes average throughput of link $l$, and $[\cdot]^{+}$
denotes $\max(0,\cdot)$.


\section{TCP Reno over A-CSMA}

\label{sec:tcp.acsma} As stated in the previous section, A-CSMA scheduling
algorithm is throughput optimal and can be easily implemented
by slightly modifying the existing L-CSMA \cite{jiang2008distributed,liu2010towards,ni2010q,rajagopalan2008distributed,chen2010markov}.
Observing the large installed base of TCP Reno, a natural question to ask before
actual deployment of A-CSMA network is how well TCP Reno can perform over A-CSMA networks.

\begin{figure*}[htbp]
\centering
\subfigure[]{
\includegraphics[scale=0.5]{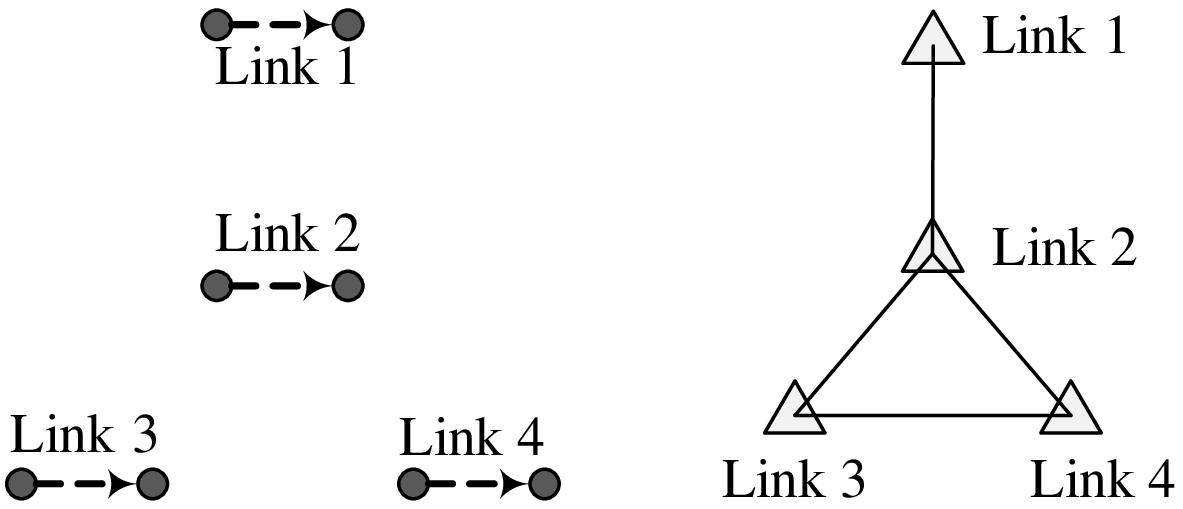}
\label{fig:simple.case}
}
\subfigure[]{
\includegraphics[scale=0.34]{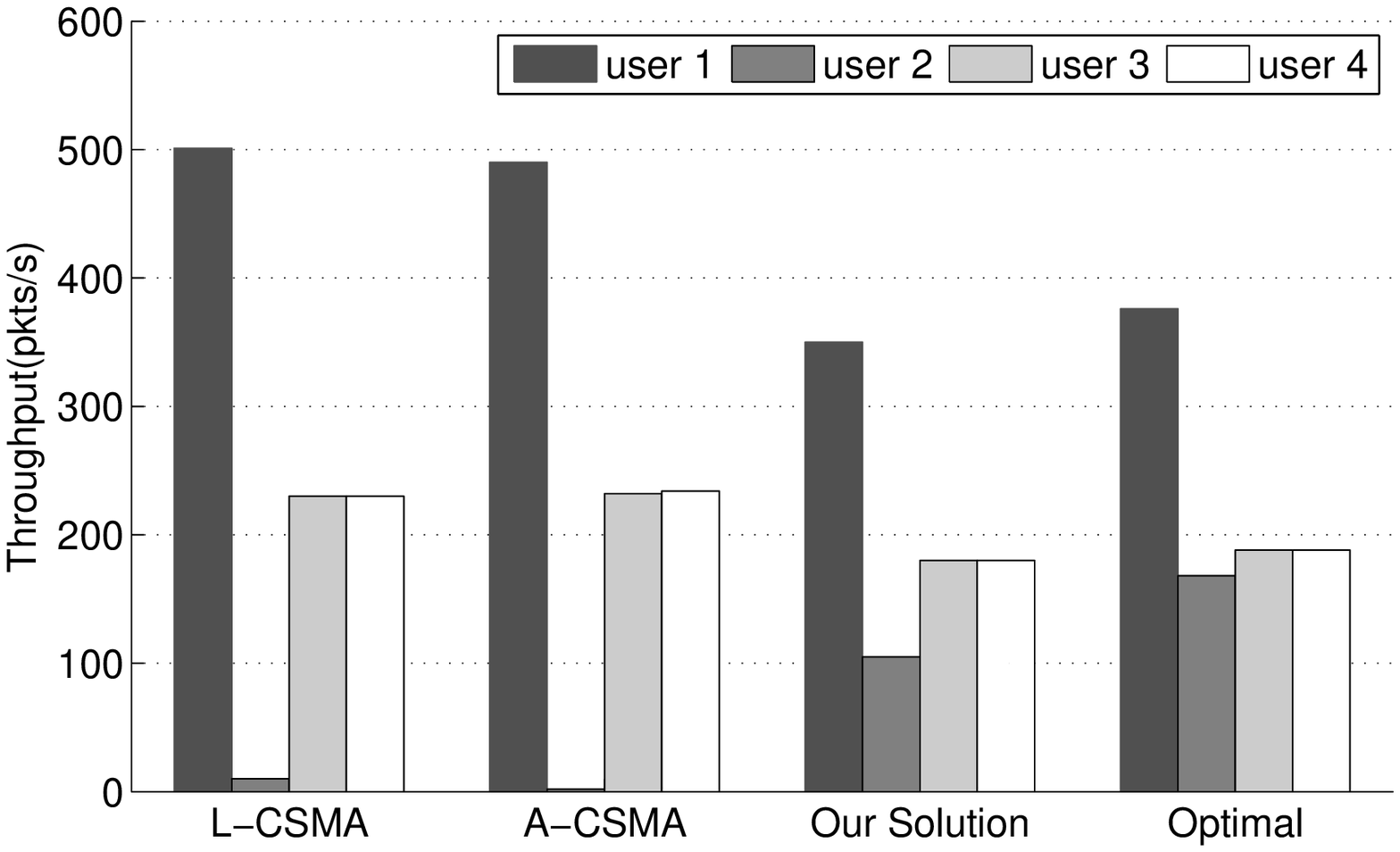}
\label{fig:tcp.acsma.throughput}
}
\subfigure[]{
\includegraphics[scale=0.20]{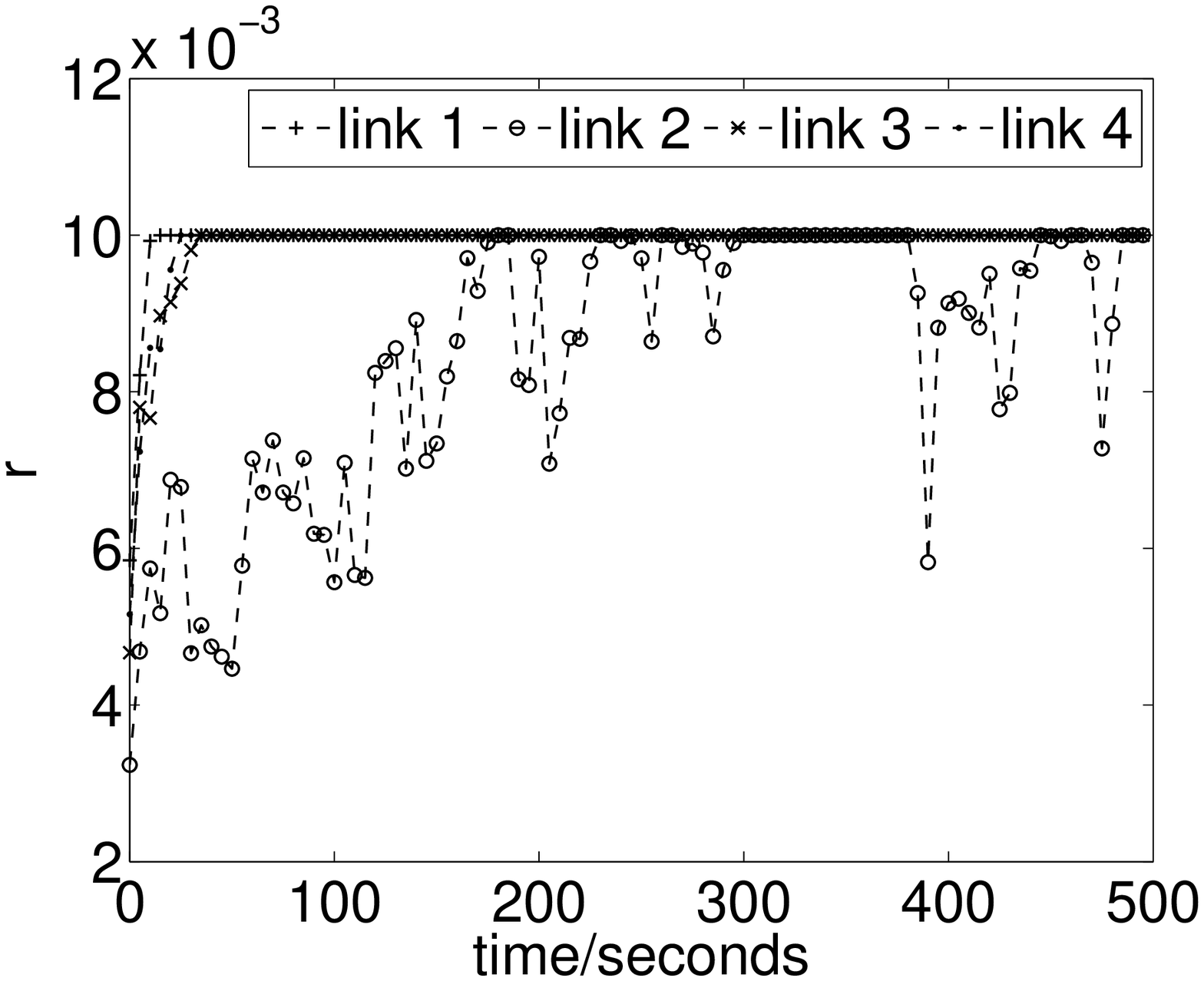}
\label{fig:result.simple.case}
}
\caption{(a) a wireless network of four links. Its conflict graph is shown on the right hand side. (b) throughput under 1) L-CSMA: NS-2 simulation of TCP Reno over L-CSMA, 2) A-CSMA: NS-2 simulation of TCP Reno over A-CSMA, 3) our solution: NS-2 simulation of multi-connection TCP Reno over A-CSMA with AQM, 4) optimal: optimal throughput computed from problem \textbf{MP}. (c) transmission aggressiveness under A-CSMA in NS-2 simulation.}
\end{figure*}

In this section, we study the performance of TCP Reno directly running
over A-CSMA network. Surprisingly, our simulation reveals that TCP Reno over A-CSMA suffers
from starvation problem. In particular, it is no better than TCP Reno over L-CSMA network\footnote{The starvation problem of TCP Reno over L-CSMA, has been extensively studied in previous work, such as \cite{wang2005throughput, liew1854back}. We give an example of starvation of TCP Reno over L-CSMA in Appendix \ref{sec:tcp.lcsma}.}.
By applying TCP Reno, the potential advantages of A-CSMA cannot be realized.

\subsection{TCP Reno Starves over A-CSMA}
We carry out NS-2 simulation to demonstrate the starvation problem
of running TCP Reno directly over A-CSMA. The network topology
is shown in Fig.~\ref{fig:simple.case}. Each link carries a single TCP Reno connection.
The simulation settings are as follows: 1) update step size and update interval for $r$ are $0.05$ and $2.0$ respectively;
2) $\beta$: weight of entropy term is $800$; 4) data rate of wireless networks is $11$Mbps; 5) TCP payload is $1000$ byte. To prevent the countdown from being too aggressive\footnote{If transmission aggressiveness are too large, $\exp(-\beta r_l)$ should be very tiny, this will result in zero countdown in digital system.}, we set the upper bound for $\br$ to be $r_{\max}=0.01$. We remark that our simulation results
hold as long as $\beta r_{\max}$ is bounded. To get a clear understanding,
we only consider the forward link (the link that transmits TCP DATA) in the analysis by letting the
backward link send back the TCP ACK with higher priority than forward link TCP DATA.
To achieve this goal, we set the backward link with a fixed maximum transmission aggressiveness $r_{max}$.
Consequently, most of the packets are buffered in the queue of the forward links.

Fig.~\ref{fig:result.simple.case} shows the throughput
achieved under TCP Reno over A-CSMA and that under TCP Reno over L-CSMA respectively. We can
see that they have nearly the same performance. In both cases, user $2$
has less throughput than other users does and therefore starves. These results show that TCP Reno over A-CSMA also suffers from severe starvation. Effectively, its performance is no better than that of TCP
Reno over L-CSMA, and the potential of A-CSMA cannot be realized by running TCP Reno directly over it. Because of the large installed base of TCP Reno, many Internet services such as HTTP and FTP that relies on TCP will suffer over A-CSMA networks.

\subsection{Observations and Explanations}

The explanation of this starvation phenomenon is as follows. Initially
all links have the same $r$ and compete the channel with the same
level of transmission aggressiveness. Under the link interference relationship, links
$1,$ $3$ and $4$ will take turns to freeze link $2$, and link
$2$ is not able to obtain its fair share. This is the same as the
origin of the starvation problem of L-CSMA discussed in Appendix \ref{sec:tcp.lcsma}.
To avoid starvation, link $2$ must be able to count down with larger
transmission aggressiveness than other links, so as to compete more
aggressively for the fair channel share against the other three (less
aggressive) links.

However, with TCP Reno running over A-CSMA, link $2$ actually obtains
a smaller transmission aggressiveness than other links, exacerbating
its starvation. We explain this as follows. TCP Reno increases its rate more slowly
when it experiences larger RTT, and vice versa. Initially, link $2$
suffers from freezing of its backoff countdown for longer time because of the transmissions of its neighbors. Consequently, TCP Reno over link $2$ experiences larger RTT
and increases its congestion window more slowly than those over other links.
In our example, the TCP congestion window size roughly equals to the
buffer length, thus link $2$'s queue increases more slowly than those
of other links. Since A-CSMA sets the transmission aggressiveness to be proportional
to the queue length, link $2$ ends up having the smallest transmission
aggressiveness among all the four links. This explanation is verified by the simulation results on transmission aggressiveness in Fig. \ref{fig:result.simple.case}.

This is a positive feedback loop. Larger transmission aggressiveness
will lead to shorter RTT. Shorter RTT makes TCP Reno increase its
rate faster, thus the larger congestion window. Larger congestion
window leads to longer queues, thus larger transmission aggressiveness.
This loop continues until the transmission aggressiveness of all the
links reach the same upper limit $r_{\max}$. With $r_{l}=r_{\max}$
for $l=1,2,3,4$, the performance of TCP Reno over A-CSMA falls back
to the performance of TCP Reno over L-CSMA, and link $2$ suffers
starvation as discussed in Appendix \ref{sec:tcp.lcsma}.

%
{}

\section{Proposed Solution}

\label{sec:solu}
In this section, we propose to use a layered approach of multi-connection
TCP Reno over A-CSMA with AQM as a practical solution to achieve fairness. This solution is provably optimal under NUM framework. Interestingly, it can be extended to wired network and multihop networks with wired and wireless links without any change. In this solution, we stick to today's TCP Reno instead of designing a new TCP for the following reasons. First, many Internet services (e.g. HTTP, FTP) currently rely on the most widely-installed TCP Reno. Second, TCP Reno has been well-tested in the Internet scale. Last, compatibility concern seems challenging and not well-understood yet \cite{tang2007equilibrium}. A new TCP needs to be carefully designed to be compatible with today's TCP Reno.

\subsection{Proposed Solution: Multi-connection TCP Reno Scheme}
\label{sec:mtcp}
We observe that TCP Reno interacts with A-CSMA through both RTT and packet loss rate. First, any loss based TCP will interact with the queue-based A-CSMA through packet loss rate. For the interaction to behave properly, we always need to employ AQM to calibrate the mapping between the link queue-length and the link loss rate. Second, it has been observed and discussed in the previous section that the reason of the flow starvation is that it suffers long RTT, which is caused by the long MAC access delay. To remove TCP bias on RTT, one way is to open multiple number of connections proportional to the RTT that TCP experiences. In this way, TCP Reno with large RTT that will open more connections to compensate for the small congestion window of each connection. As a result, there are more outstanding packets filling in the queue of the link. This makes A-CSMA allocate more airtime to the link, reducing its MAC accessing delay of the link and thus the RTT of the flow.

Inspired by the above observations, we propose to use multi-connection TCP Reno to remove the RTT interaction between TCP Reno and A-CSMA, and use AQM to calibrate their packet loss interaction.

In our solution, every user $s$ monitors the round trip time $T_s$, and opens $n_s = kT_s$ (where $k$ is a constant) TCP Reno connections to remove the RTT bias.
We remark that now the user rate $x_s$ is the aggregate rate of $n_s$ TCP Reno connections. We present the overall scheme as the following dynamic system:
\begin{numcases}{}
    n_s = k T_s, \;\forall s\in \mathcal{S}; \label{eq:ns}\\
   \dot{x_s}= \frac{x_{s}^{2}}{2n_s}\left(\frac{2n^2_s}{T_{s}^{2}x_{s}^{2}}-\sum_{l\in s}r_{l}\right)_{x_s}^+, \;\forall s\in \mathcal{S}; \label{eq:mtcp} \\
   \dot{r_l}= \alpha \big[\sum_{s:l\in s}x_s - \sum_{i:l\in i}\tau_i (\br) \big]_{r_l}^+, \; \forall l\in L, \label{eq:r}
\end{numcases}
where \begin{equation}
\tau_i (\br) = \frac{\exp\left( \sum_{l\in i} \beta r_l\right)}{\sum_{i^{\prime}\in \mathcal{I}}\exp\left( \sum_{l\in i^{\prime}} \beta r_l\right)}, \; \forall i\in \mathcal{I}. \label{eq:tau}\end{equation}
We remark that how $T_s$ is updated is irrelevant to our solution as we always compensate for its impact by opening $n_s$ connections. The multi-connection TCP Reno dynamic equation can be expressed in the form of (\ref{eq:mtcp})\footnote{Recall that the single connection TCP Reno congestion window dynamic equation is,
\begin{equation}
    \dot{W_s} = \frac{1}{T_s} -\frac{W_s^2}{2T_s} (\sum_{l\in s} p_l)\label{eq:w},
\end{equation}
where $T_s$ is the round trip time and $\sum_{l\in s} p_l$ is the end-to-end packet loss rate experienced by the TCP Reno. If flow $s$ opens $n_s$ number of TCP connection, the flow rate $x_s$ can be expressed as \begin{equation}x_s = \frac{n_s W_s}{T_s}.\label{eq:xn}\end{equation}
From (\ref{eq:w}) and (\ref{eq:xn}), we now have
\begin{equation}
   \dot{x}_s = \frac{x_s^2}{2 n_s}( \frac{2 n_s^2}{T_s^2x_s^2} - \sum_{l \in s} p_l ) \label{eq:xs}.
\end{equation}
} \cite{chen2006flow} when the end-to-end price (e.g., packet loss rate) is $\sum_{l\in s} r_l$. This link price can be fed back by packet loss based technique like Active Queue Management (AQM). In this paper, we adopt an AQM policy in which each link drops or ECN-marks packet with probability equal to $r_l$. We remark that the independent sets distribution satisfies (\ref{eq:tau}) as long as each link applies CSMA mechanism and backoff with mean equal to $\exp(-\beta r_l)$. Dynamic equations (\ref{eq:r}) are exact to A-CSMA algorithm introduced in Section \ref{sec:acsma} and $r_l$ is proportional to queue length.

The above dynamic system achieves certain system utility. With the time-scale separation assumption, it is stable and converges to the unique equilibrium. We summarize the result as the following theorem.

\begin{theorem}\label{the:sol}
\begin{enumerate}
    \item Equilibrium of dynamic system in (\ref{eq:ns})-(\ref{eq:r}) solves the following Network Utility Maximization problem\footnote{This formulation is similar to rate control over TDMA network except that there is an additional entropy term in the objective function. As shown in \cite{jiang2008distributed,chen2010markov}, this leads to distributed implementation by using A-CSMA.}:
    \begin{eqnarray}
    \mbox{\bf {MP}:} & \max_{\bx\geq0,\btau\geq0} & \sum_{s} -\frac{1}{x_s}-\frac{1}{\beta}\sum_{i\in\mathcal{I}}\tau_{i}\log\tau_{i}\nonumber \\
    & \mbox{s.t.} & \sum_{s:l\in s}x_{s}\leq\sum_{i:l\in i}\tau_{i},\;\;\forall l\in E,\label{eq:flow}\\
    &  & \sum_{i\in\mathcal{I}}\tau_{i}=1,\label{eq:is}
    \end{eqnarray}
    \item With the time-scale separation, i.e., the A-CSMA Markov chain converges to its stationary distribution instantaneously, the dynamic system (\ref{eq:ns})-(\ref{eq:r}) globally asymptotically converges to the unique optimal solution of problem \textbf{MP}.
\end{enumerate}
\end{theorem}
Remark: at the equilibrium, the utility function $-\frac{1}{x_s}$ guarantee an $\alpha$-fairness among users with $\alpha =2$ \cite{mo2000fair}. One of its implications is that \emph{no user will
starve at the optimal solution at the equilibrium}. This provides theoretical justification that our proposed solution will effectively address the TCP-Reno over A-CSMA starvation problem.
The proof of Theorem \ref{the:sol} 1) is given in Appendix \ref{sec:proof}.

Theoretically, under the fluid model analysis, the proof of convergence can still be established under some mild conditions on parameters. The ideas are very similar to those expounded in \cite{jiang2009convergence,shao2010cross}. We skip the proof here due to space limitation.

Without such time-scale separation assumption, the dynamic system turns into a stochastic dynamic system, where the link rate measured by link $l$ does not satisfy equations (\ref{eq:tau}). Under small step size and update interval, the resulted stochastic dynamic system is shown to converge to a bounded neighborhood of the same optimal solutions with probability one \cite{chen2010markov,shao2010cross}.

We validate the convergence of the algorithm in packet-level simulations later in Section \ref{sec:simul}.

\subsection{Implementation}
\label{sec:imp}
Inspired by the observations from the dynamic systems in (\ref{eq:ns})-(\ref{eq:r}). We can solve the problem by running multi-connection TCP Reno over A-CSMA with AQM. We stress multi-connection is important in our solution in that it removes the RTT interaction with A-CSMA. Without removing such RTT bias on RTT, TCP Reno over A-CSMA with AQM also results in starvation problem. Interested readers can refer to our Appendix \ref{sec:acsma.aqm} for detailed results and analysis.

The multi-connection layer can be implemented in two ways, without modification to the transport layer. One method is to insert an intermediate layer between the application layer and transport layer, called Multi-connection API. It provides the universal API to the application layer, and functions to maintain $k T_s$ TCP Reno connections from monitored RTT. This solution requires no modification to TCP/IP stack and is compatible with today's applications (like FTP, HTTP, etc.).

The other implementation method is to rewrite the applications. The application keeps monitoring RTT and opens multiple sockets. To obtain the RTT, we can simply set up a UDP connection to measure the RTT~\cite{chen2006multiple,chen2006flow}.

\begin{figure}[H]
\centering
  \includegraphics[scale=0.6]{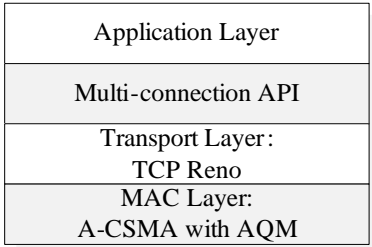}
\caption{Scheme Diagram}
\label{fig:ps.dia}
\end{figure}

We prefer the first implementation in which we insert an multi-connection API in between. This implementation does not require any modification to today's applications and hence simplifies programming. From the structure of the dynamic system in (\ref{eq:ns})-(\ref{eq:r}), it can be implemented in a layered manner as follows, with the diagram shown in Fig~\ref{fig:ps.dia}:
\begin{itemize}
\item in the MAC layer, we run A-CSMA to schedule the link transmissions. This is directly obtained from dynamic equations (\ref{eq:r}). A-CSMA maintains transmission aggressiveness vector $\br$ to be proportional to the lengths of links' queue.

\item we run AQM at each link that drops or marks packets with a probability proportional to $\br$, and thus proportional to the queue lengths. In this way, the prices of using individual links can be fed back to end users via packet losses or ECN marks. We remark that AQM is required if we want to run loss based TCP (e.g. Reno) over A-CSMA.

\item in the transport layer: we perform rate control based on sum of the prices of individual links, which is fed back in a form of packet losses or ECN marks. This result is directly obtained from equations (\ref{eq:mtcp}). When loss ratio of link $l$ is $r_l$, the total loss ratio that TCP Reno sees is $1-\prod_{l \in s}(1-r_l) \approx \sum_{l \in s}r_l$, when $r_l$ is small.

\item multi-connection layer, we maintain $n_s = kT_S$ connections to remove the RTT bias.
\end{itemize}
We remark that AQM is minimally required in order to run loss-based TCP (e.g. TCP Reno) over A-CSMA. The pseudo code of A-CSMA with AQM is shown in Alg.~\ref{alg:A1}.

\begin{algorithm}[hbt]\caption{: Pseudo code of A-CSMA with AQM.}
\begin{algorithmic}[1]
\label{alg:A1}
    \STATE{// Updating the transmission aggressiveness:}
           \IF{at the end of time unit $t-1$}
             \STATE{Transmission aggressiveness $r_l \leftarrow \alpha \cdot Q_l(t-1)$.}
           \ENDIF
    \STATE{// Setting the backoff window:}
           \IF{before transmitting a packet with size $PKT$}
              \STATE{Sets the backoff window following exponential distribution with mean equal to $PKT* \exp(-\beta r_l)$.}
           \ENDIF
    \STATE{// Adjusting the queue length:}
           \IF{a packet is served}
              \STATE{Queue length $Q_l(t) \leftarrow max(Q_l(t) - 1,0)$.}
           \ENDIF
           \IF{a new packet is inserted into queue}
              \STATE{(a) Discards tail packet with probability $\min(r_l,1)$.}
              \STATE{(b) $Q_l(t) \leftarrow Q_l(t) + 1$.}
           \ENDIF
           \IF{no packet to serve in the queue}
              \STATE{Generates a dummy packet.}
           \ENDIF
\end{algorithmic}
\end{algorithm}

In practice, user $s$ can only open an integer number of connections. By setting $n_s=\mbox{int}(k T_s)$, user $s$ can react to the change in $T_s$ in fine granularity if $k$ is large, and in coarse granularity for small $k$. While large $k$ is preferred in theory, in practice user $s$ may need to maintain a large number of connections if $k$ is too large. Large $k$ also make the $n_s$ adjustment more responsive to the changes in $T_s$, so it induces more overhead due to frequently opening and closing TCP Reno connections.

Packet loss ratio should be far less than one, otherwise it will waste considerable bandwidth for dropping large number of packets. Besides, the total loss ratio $1-\prod_{l \in s}(1-r_l)$ does not hold when $r_l$ is large. We can scale $r_l$ so as to make it much smaller than one.

\subsection{Discussion}\label{sec:discu}
We discuss the following two questions in this subsection: 1) By using multi-connection TCP Reno, can we achieve arbitrary utility function? 2) Is our solution applicable to wired network as well as multihop networks with wired and wireless links?

\subsubsection{Achieve Arbitrary Utility}
For the first question, our answer is yes. Let $U_s(x_s)$ be arbitrary utility function of problem \textbf{MP}. Consider our solution in (\ref{eq:ns})-(\ref{eq:r}) but replacing the updating equation of $n_s$ as follows:
\begin{equation}
    n_s = k\sqrt{\frac{U_s'(x_s)T_s^2 x_s^2 }{2}}.
\end{equation}
It is not difficult to verify that the equilibrium of the modified solution solves problem \textbf{MP} with the specified utility function $U_s(x_s)$.

In particular, our proposed solution can be understood as a special case of the above approach with the specified utility function being $U_s(x_s) = -\frac{1}{x_s}$.

\subsubsection{Extension to Networks with Both Wired and Wireless links}
Our answer to the second question is also yes. We consider a multihop network composed of wired and wireless links. Let $l$ denote the wireless link and $w$ denote the wired link. Our dynamic system to solve the extension problem of networks composed of wired and wireless links is as follows:
\begin{numcases}{}
    n_s = k T_s, \;\forall s\in \mathcal{S}; \label{eq:ns2}\\
   \dot{x_s}= \frac{x_{s}^{2}}{2n_s}\left(\frac{2n^2_s}{T_{s}^{2}x_{s}^{2}}-\sum_{l\in s}r_{l} - \sum_{w \in s} p_w\right)_{x_s}^+, \;\forall s\in \mathcal{S}; \label{eq:mtcp2} \\
   \dot{r_l}= \alpha \big[\sum_{s:l\in s}x_s - \sum_{i:l\in i}\tau_i (\br) \big]_{r_l}^+, \; \forall l\in E^{\prime}, \label{eq:r2}
\end{numcases}
where \begin{eqnarray}
   \tau_i (\br) = \frac{\exp\left( \sum_{l\in i} \beta r_l\right)}{\sum_{i^{\prime}\in \mathcal{I}}\exp\left( \sum_{l\in i^{\prime}} \beta r_l\right)}, \; \forall i\in \mathcal{I}. \label{eq:tau2}\\
    p_w = \frac{(\sum_{s':w\in s'} x_{s'} - C_w)^+}{\sum_{s':w\in s'} x_{s'}}, \; \forall w\in E_w, \label{eq:pw}\end{eqnarray}
and $E_w$ denotes the set of wired links, $E_l$ denotes the set of wireless links, $C_w$ stands for the capacity of wired link $w$, and $p_w$ is the packet loss ratio of wired link (when it applies drop-tail queue).

\begin{theorem}\label{the:ext}
Equilibrium of dynamic system in (\ref{eq:ns2})-(\ref{eq:r2}) solves the following Network Utility Maximization problem:
\begin{eqnarray}\textbf{EP:}
    & \max & \sum_s -\frac{1}{x_s}- \frac{1}{\beta}\sum_{i\in\mathcal{I}}\tau_{i}\log\tau_{i} \nonumber\\
        && - \sum_{w\in E_w} \int_0^{\sum_{s':w\in s'} x_{s'}} \frac{(y-C_w)^+}{y}dy \\
    & s.t. & \sum_{s:l\in s} x_s \leq \sum_{i:l\in i} \tau_i, \;\;\forall l\in E^{\prime}, \\
          && \sum_{i\in \mathcal{I}} \tau_i = 1, \\
          && \bx \geq 0, \btau \geq 0,
\end{eqnarray}
    where the utility function is $U_s(x_s) = -\frac{1}{x_s},\;\forall s \in \mathcal{S}$ and $\int_0^{\sum_{s':w\in s'} x_{s'}} \frac{(y-C_w)^+}{y}dy$ is the penalty function which penalizes the arrival rate for exceeding the wired link capacity $c_w$.
\end{theorem}
This theorem can be proved using similar method of the proof of Theorem \ref{the:sol}.

From the above extension dynamic system, we have the following observations: 1) in MAC layer, wireless links perform A-CSMA with AQM; 2) link dropping ratio at wireless link is exact equal to packet dropping probability at today's routers (with drop-tail queue). Therefore, our solution discussed in Section \ref{sec:imp} can be directly extended to networks with wired and wireless links without modification to any network components.

\section{Simulations}

\label{sec:simul} We evaluate the performance of multi-connection TCP Reno over A-CSMA stated in Section \ref{sec:mtcp} with NS-2 simulation. A-CSMA with AQM is implemented following the description of Algorithm \ref{alg:A1} by modifying  the IEEE 802.11 module. Unless specified otherwise, the simulation parameters are the same. All simulations are conducted with zero channel losses, so as to give us a clear observation. By default, 1) $k$: the coefficient for $n_s$, is $10$; 2) update step size $\alpha$ and update interval for $r$ are $0.05$ and $2.0$second, respectively; 3) the weight of entropy term $\beta = 2000$; 4) data rate is $11$Mbps; 5) carrier sensing range and transmission range are $800$m and $250$m respectively to avoid hidden-node problem. We do not wish to vary the number of TCP Reno connection too dynamically. In the simulation we update the connection number every $5$ seconds. In the following subsections, we will investigate our solution under three different scenarios: single-hop networks, multihop pure wireless networks and multihop networks with wireless links and wired links.


\subsection{Single-hop Networks Scenario}
In this part, we consider single-hop wireless link data transmission. We run TCP Reno over L-CSMA, TCP Reno over A-CSMA, and Multi-connection TCP Reno over A-CSMA with AQM on the given topologies. We measure the throughput achieved by each user in different schemes, and compare these throughput with the optimal value. The optimal throughput is computed by solving the utility maximization problem \textbf{MP}.

\begin{figure*}[htbp]
   \centering
   \subfigure[Topology $a$]{
   \includegraphics[scale=0.6]{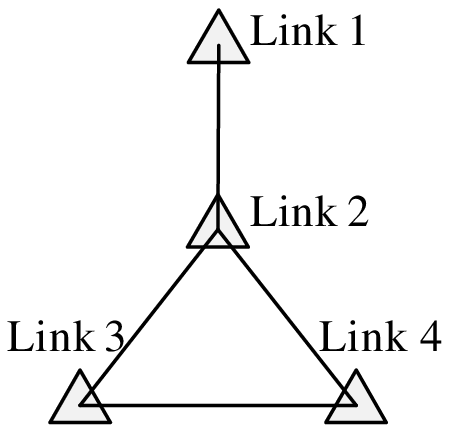}
   }
   \hspace{1in}
   \subfigure[Topology $b$]{
   \includegraphics[scale=0.6]{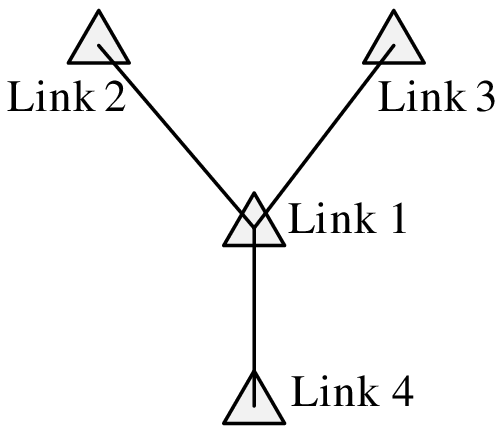}
   }
   \hspace{1in}
   \subfigure[Topology $c$]{
   \includegraphics[scale=0.6]{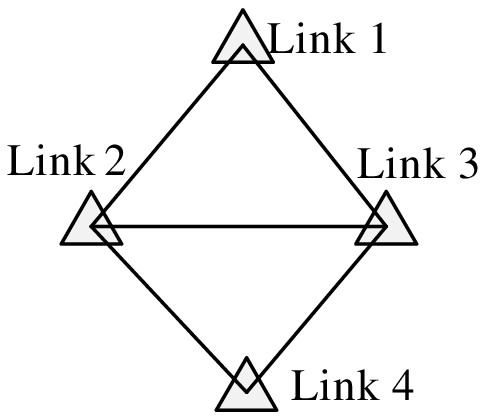}
   }
   \caption{Link Conflict Graphs of single-hop network topologies.}
   \label{fig:sh.cg}
\end{figure*}

\begin{figure*}[htb]
   \centering
   \subfigure[Topology $a$]{
   \includegraphics[scale=0.28]{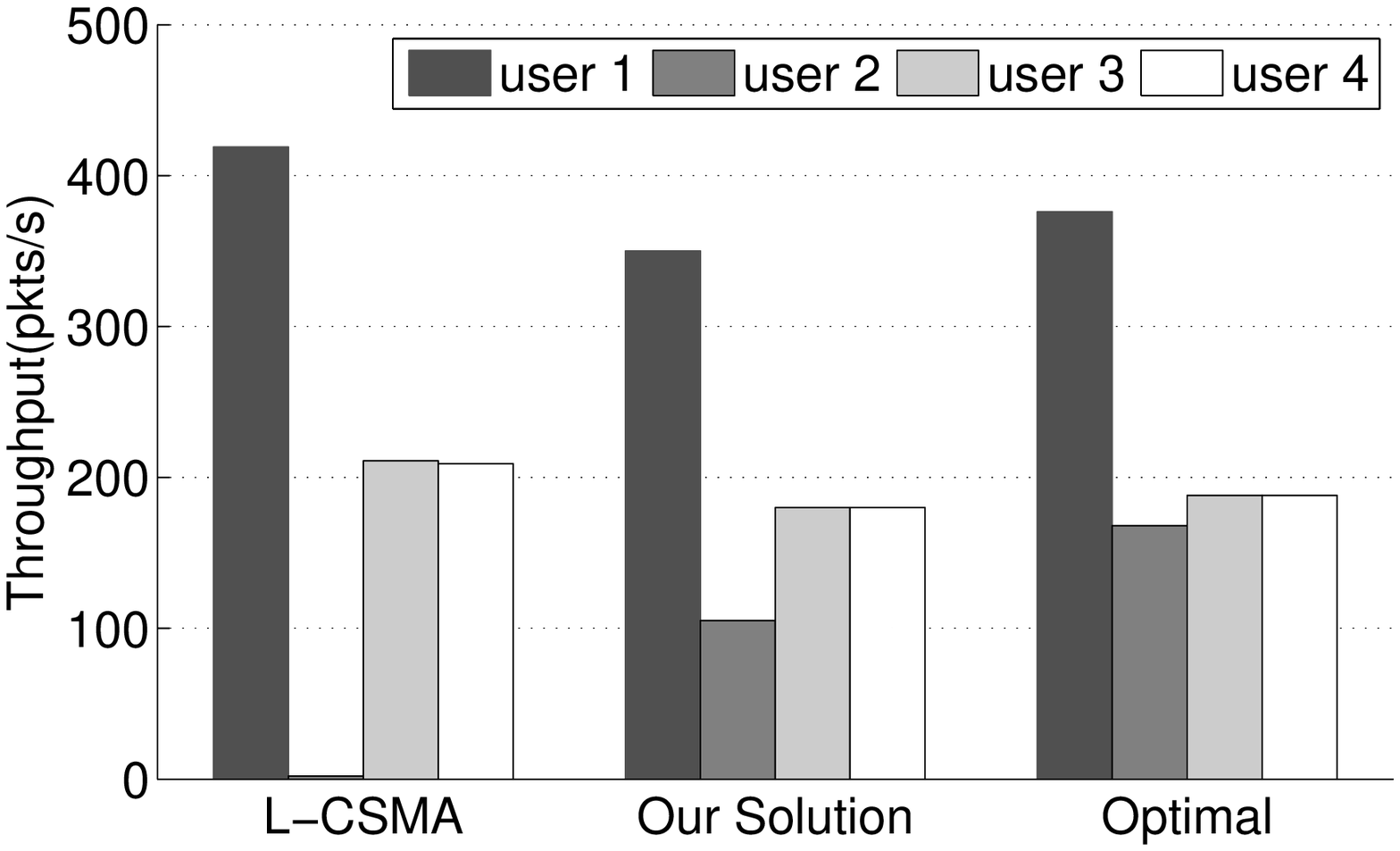}
   \label{fig:thr.s}
   }
   \subfigure[Topology $b$]{
   \includegraphics[scale=0.28]{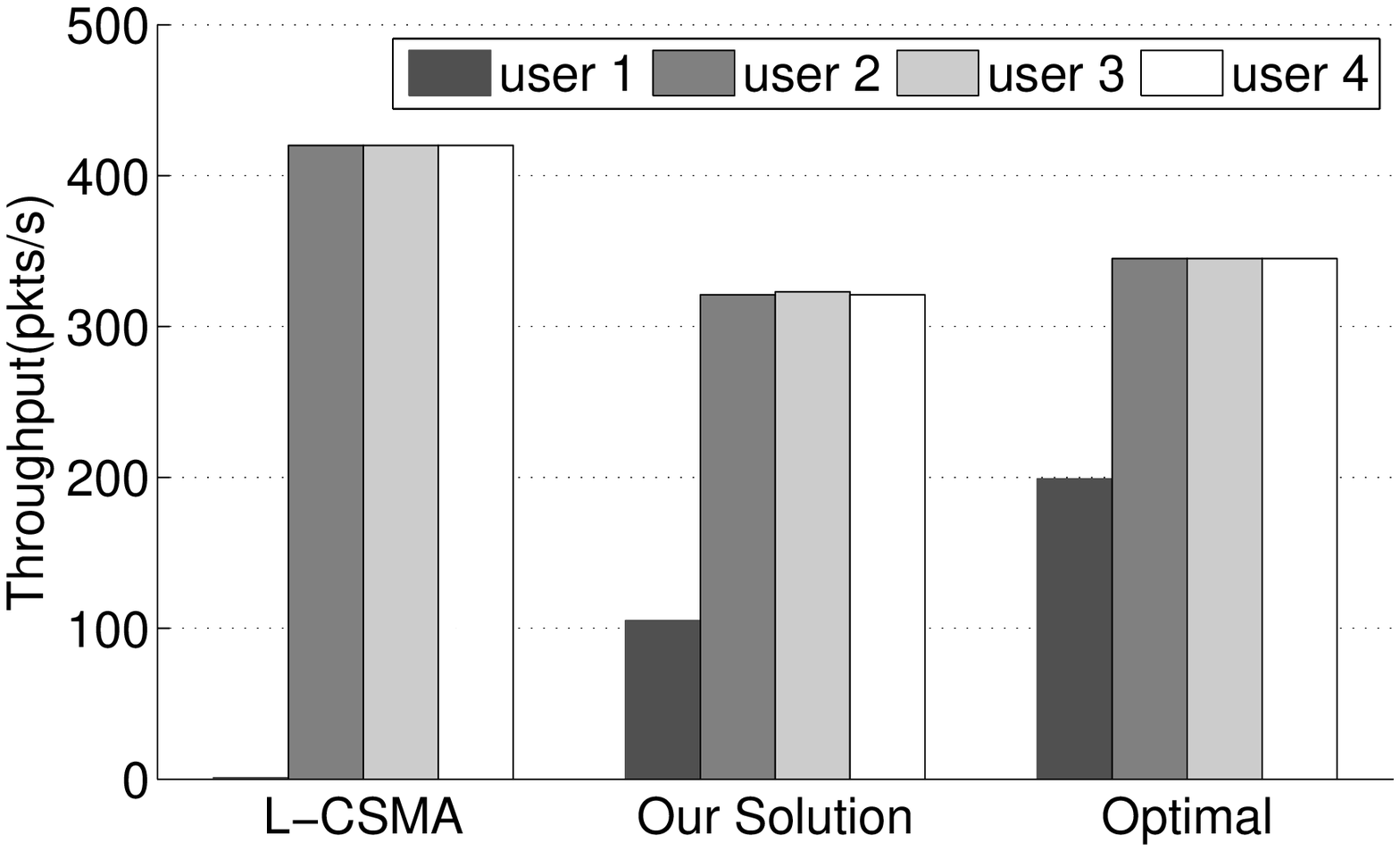}
   \label{fig:thr.y}
   }
   \subfigure[Topology $c$]{
   \includegraphics[scale=0.28]{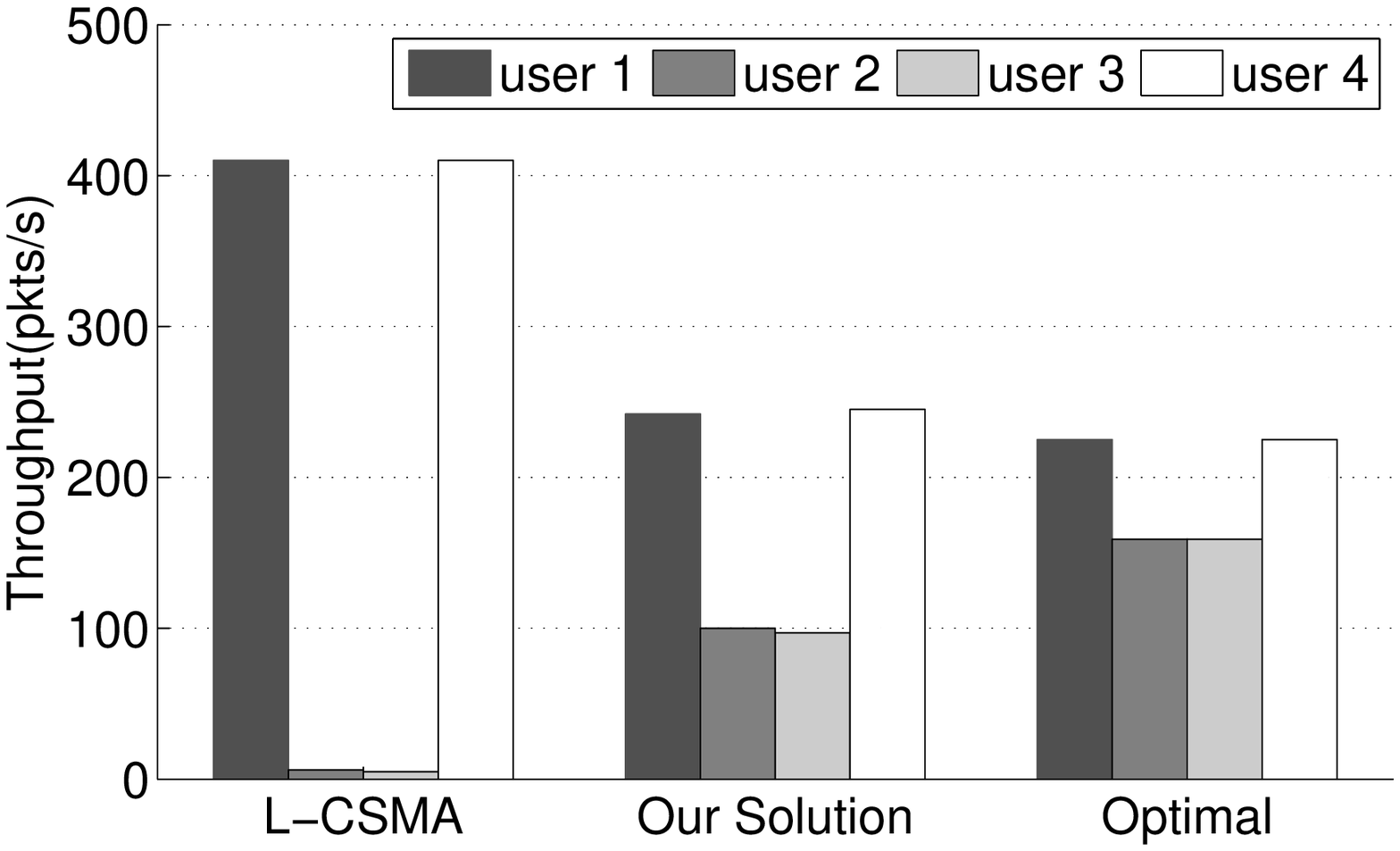}
   \label{fig:thr.dia}
   }
   \caption{Throughput under 1) L-CSMA: NS-2 simulation of TCP Reno over L-CSMA, 2) our solution: NS-2 simulation of multi-connection TCP Reno over A-CSMA with AQM and 3) optimal throughput computed from problem \textbf{MP}.}
   \label{fig:thr.sh}
\end{figure*}

To understand the performance of multi-connection TCP Reno over A-CSMA with AQM, we examine three different topologies with conflict graphs shown in Fig.~\ref{fig:sh.cg}. All topologies consist of four links, with each link carrying a single user data flow. Each of these three topologies has the starvation problem when TCP Reno is running over L-CSMA networks, which has been studied in \cite{liew1854back}. In topology a, which has been studied in previous sections, TCP flow running over link $2$ starves. In topology b, link $1$ starves and links $2$, $3$ and $4$ obtain the whole portion of channel access. In topology c, links $2$ and $3$ starve, link $1$ and link $4$ get the half of the channel bandwidth. This result can also be seen from the NS-2 simulation result in Fig.~\ref{fig:thr.sh}.

We plot the throughput obtained by different schemes in Fig.~\ref{fig:thr.sh}. Our multi-connection TCP Reno scheme guarantees the contending users in wireless network fairly share the channel. We define utility gap $\Delta U = \sum_s (U(x_s) -U(x^*_s))$ as the difference between the system achieves and the optimal utility. The closer $\Delta U$ to zero, the closer the system to the optimal. We give the utility gap under different scheme in all topologes in Table. \ref{tab:sh.utility}. This proves that our solution is capable of realizing the benefit of A-CSMA networks and achieving fair rate allocation in wireless network without any modification to TCP/IP stack in a fully distributed way. We observe that there is an optimality gap between our solution and optimal. Count down wastes some bandwidth, therefore, the performance is surely lower than optimal. The other reason is sharp oscillations of RTT. In our analysis, we assume that RTT is equal to average queueing delay plus propagation delay, which changes sharply in practice.

\begin{table}[htb]
\centering
\caption{Utility gap $\Delta U$ under different schemes}\label{tab:sh.utility}
  \begin{tabular}
  [c]{|c|c|c|c|c|c|} \hline
                          & Topo. $a$  & Topo. $b$ & Topo. $c$ &Topo. $d$ &Topo. $e$  \\ \hline \hline
    \textbf{L-CSMA}       & -268.5  &   -541.4 &  -190.8  &  -538.3 &  -1.8 \\ \hline
    \textbf{Our Solution} & -3.9    &   -2.8   &  -3.8    &   -3.5 &   -0.9\\ \hline
  \end{tabular}
\end{table}

\begin{figure}[htbp]
   \centering
   \subfigure[Transmission Aggressiveness]{
   \includegraphics[scale=0.20]{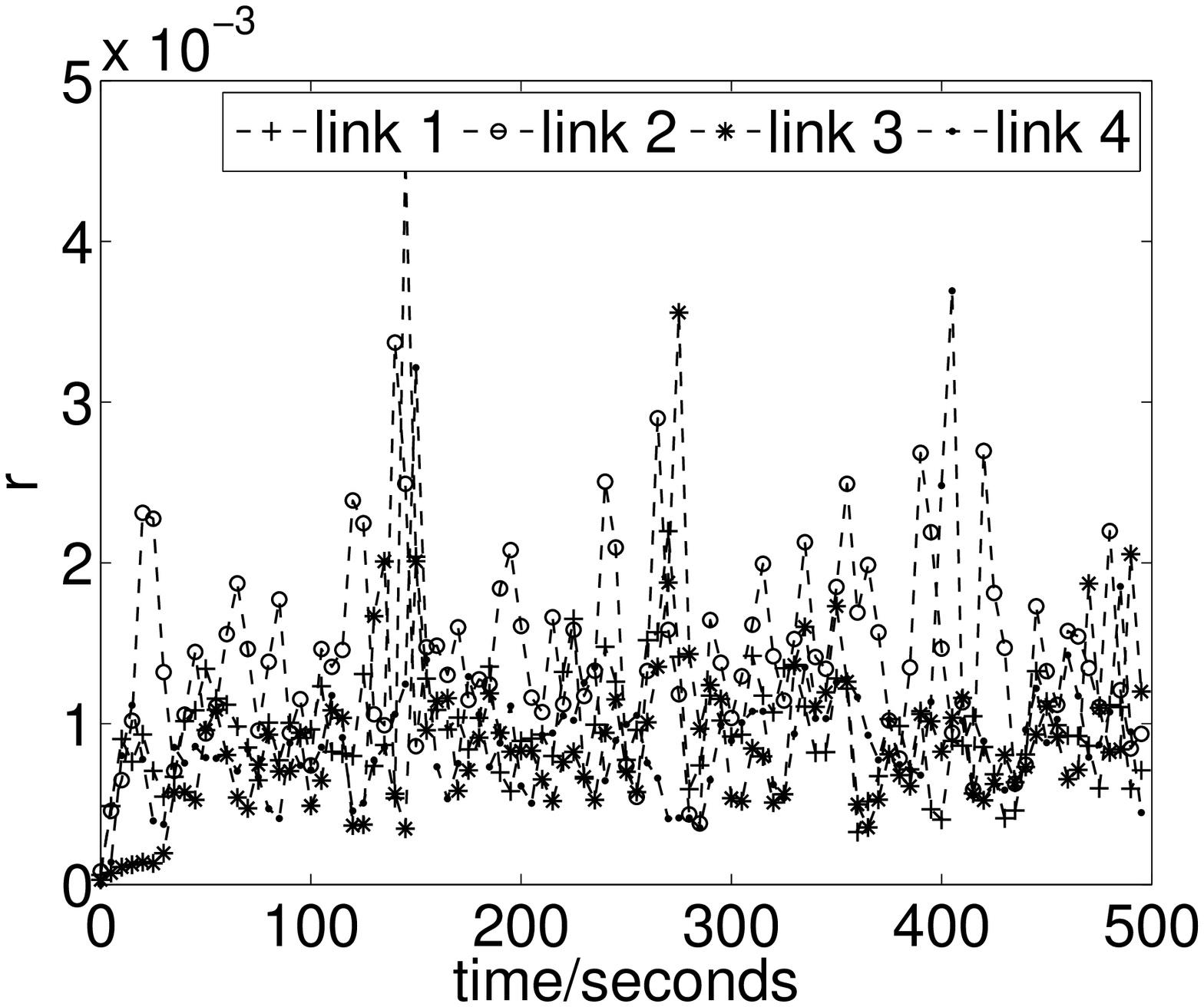}
   \label{fig:ta_ss}
   }
   \subfigure[Number of connections]{
   \includegraphics[scale=0.20]{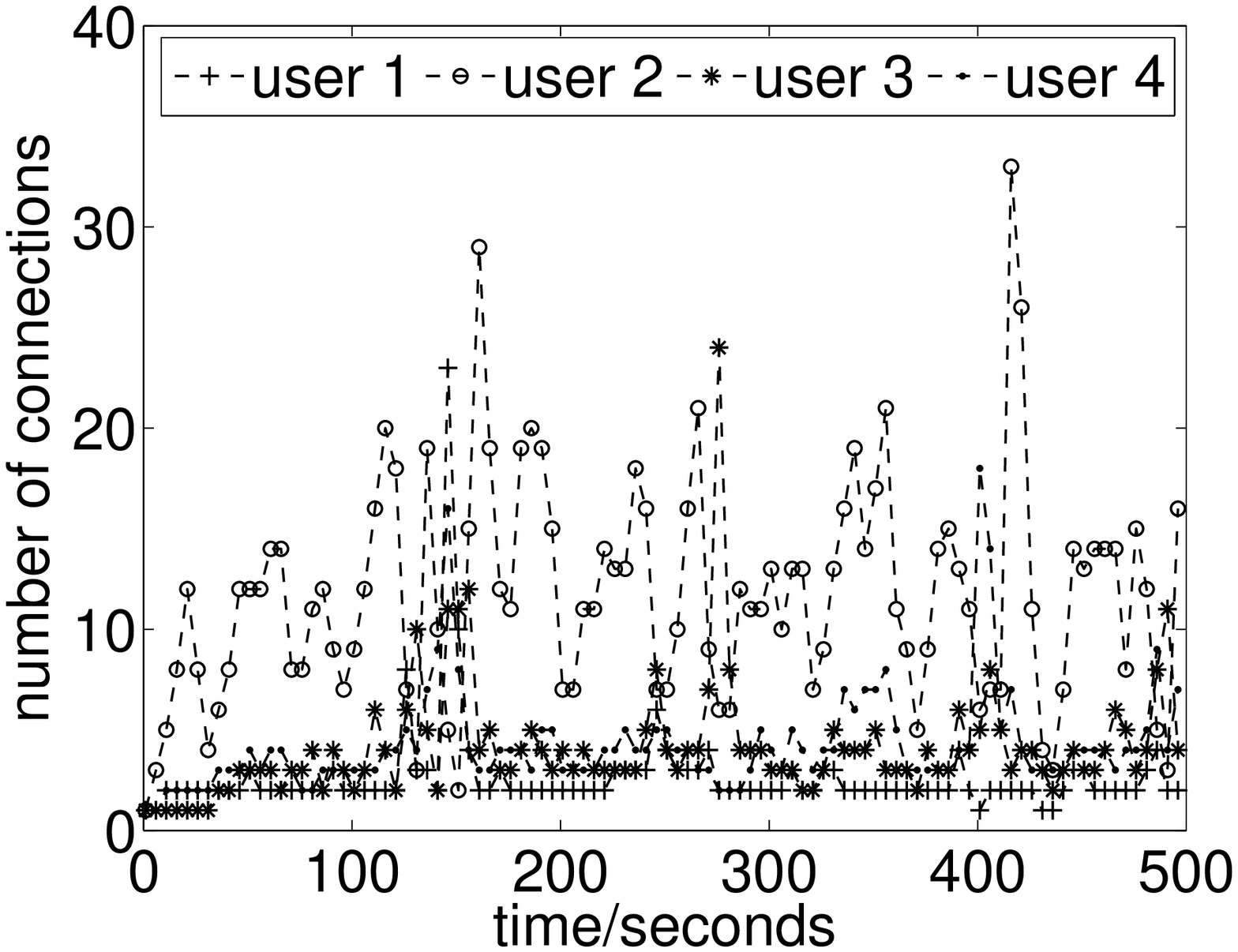}
   \label{fig:cw_ss}
   }
   \caption{Simulation results for topology $a$ under multi-connection TCP Reno over A-CSMA with AQM: curves of transmission aggressiveness and number of connections.}
   \label{fig:sh.plot}
\end{figure}

Fig.~\ref{fig:sh.plot} shows the evolution of transmission aggressiveness $r$ and number of connection $n_s$ of topology a. We observe that by applying our solution, TA $\br$ and the number of connections $n_s$ are stable. Note that dropping probability $r$ of each link, around $0.001\sim 0.003$, is very small, so it does not waste too much bandwidth by dropping large number of packets. The number of connection $n_s$ link $2$ opens is around $15$ in the simulation. The number of connections can be reduced by using a small $k$. Larger $k$ ensures finer granularity in adjusting $n_s$, thus more responsive to the change in $T_s$, and therefore better system performance, but at the cost of consuming more device resources. In reality, both link backoff procedure and transmission delay affects RTT, result in very dynamic in RTT. This is why we see sharp oscillation of $n_s$ in Fig.~\ref{fig:sh.plot}, because $n_s$ is tuned to remove RTT bias.  We can smooth the $n_s$ by take the average of the current value and last value, but at the cost of not responsive to the change in $T_s$.

The results indicate that our multi-connection TCP Reno over A-CSMA with AQM scheme achieves the fair and efficient rate allocation among different users in the single hop scenario.

\subsection{Multihop Networks Scenario}
\label{sec:mh}
\begin{figure*}[htbp]
\centering
\subfigure[]{
   \includegraphics[scale=0.4]{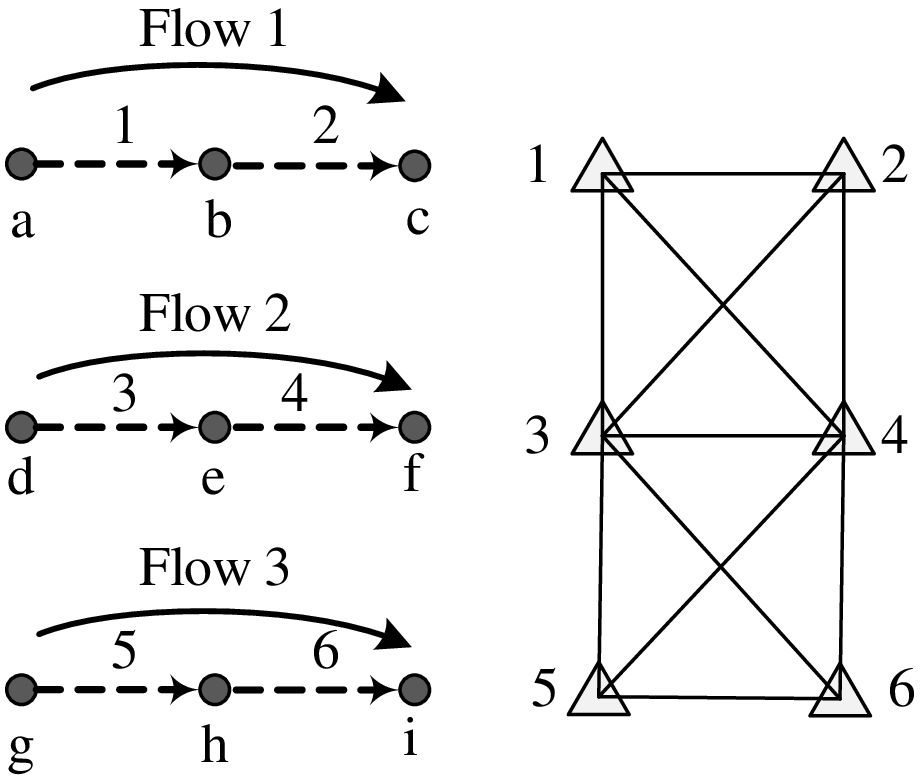}
   \label{fig:to.mh}
   }
\subfigure[]{
\includegraphics[scale=0.28]{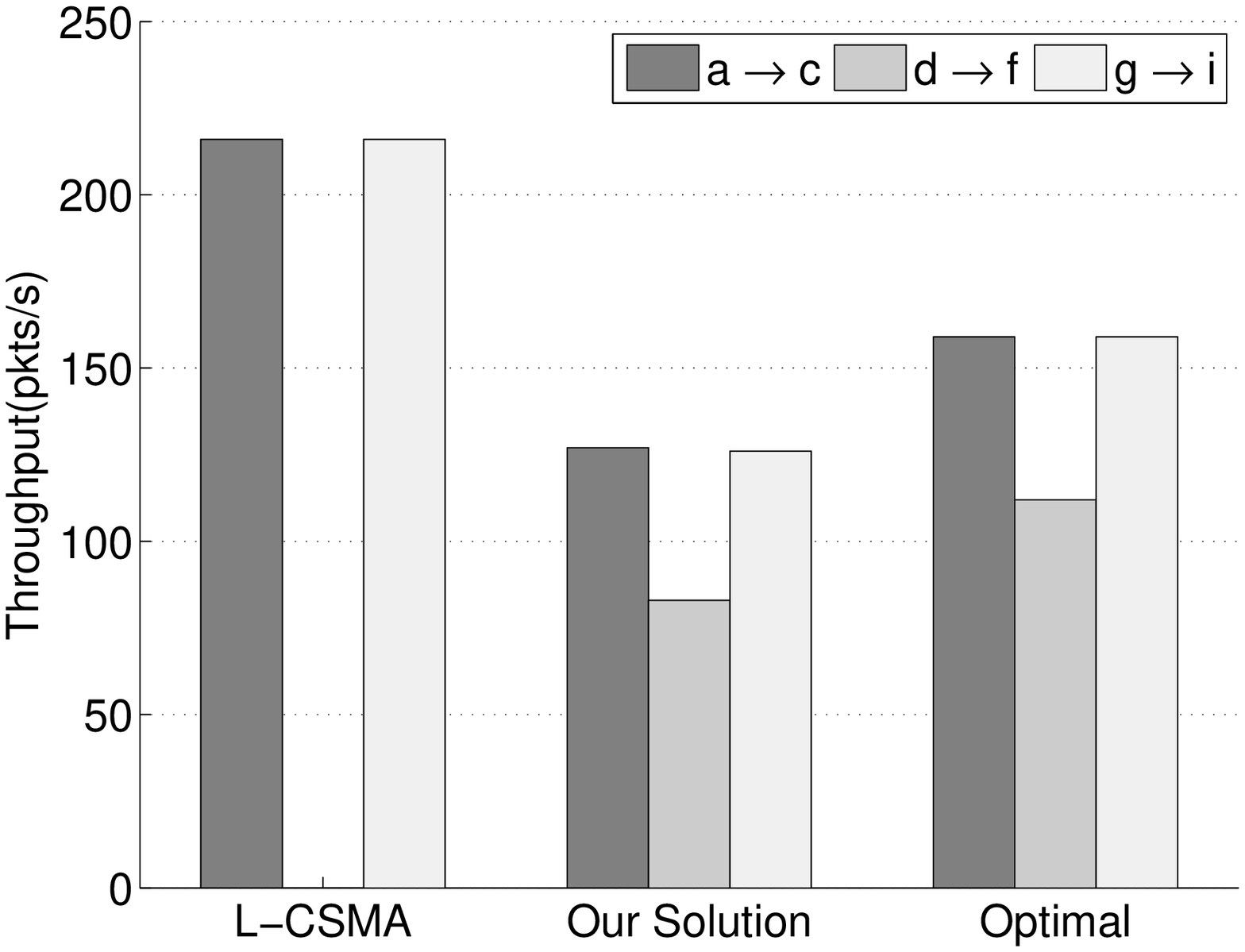}
\label{fig:thr.mh}
}
\subfigure[]{
   \includegraphics[scale=0.20]{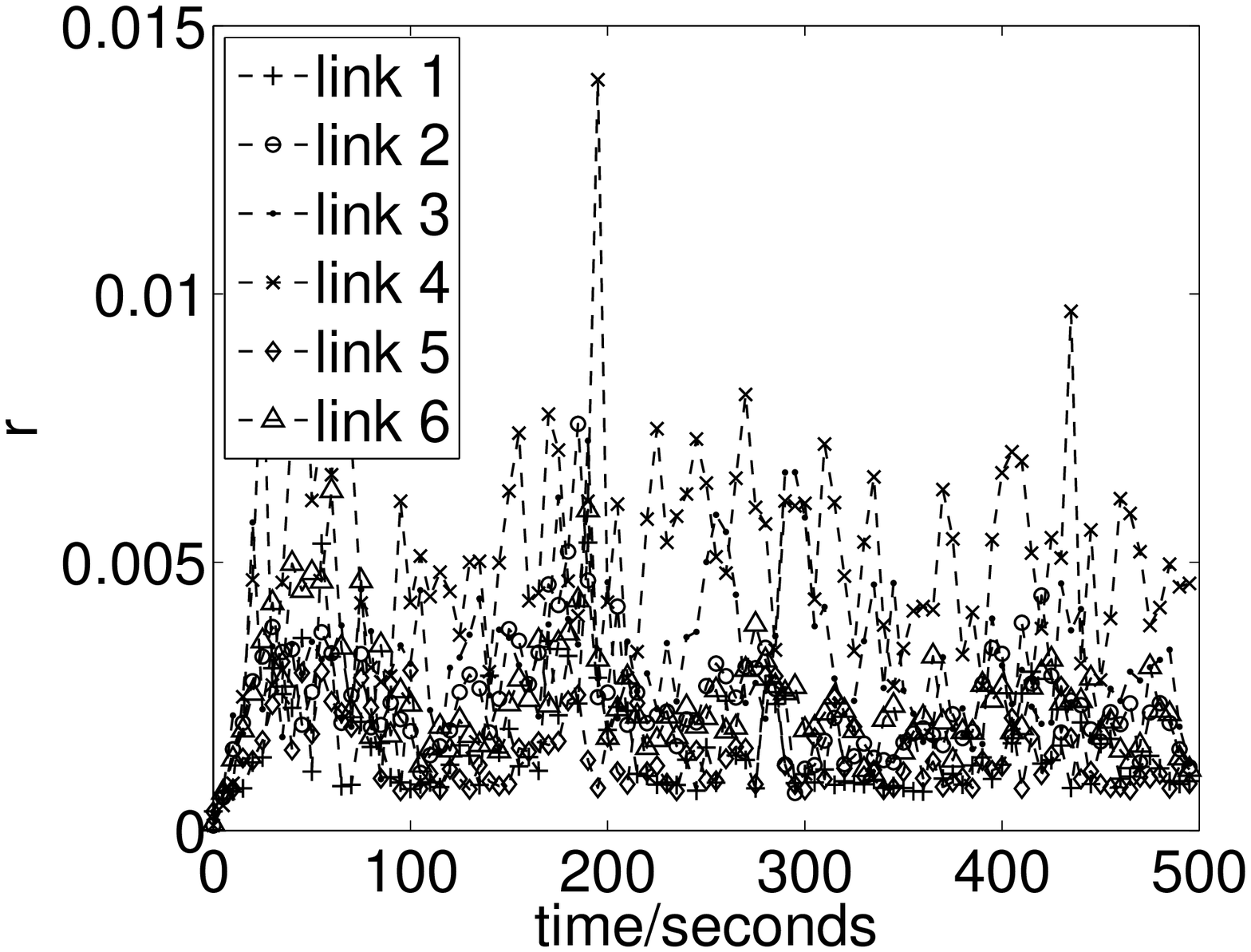}
   \label{fig:ta.mh}
   }
\subfigure[]{
   \includegraphics[scale=0.20]{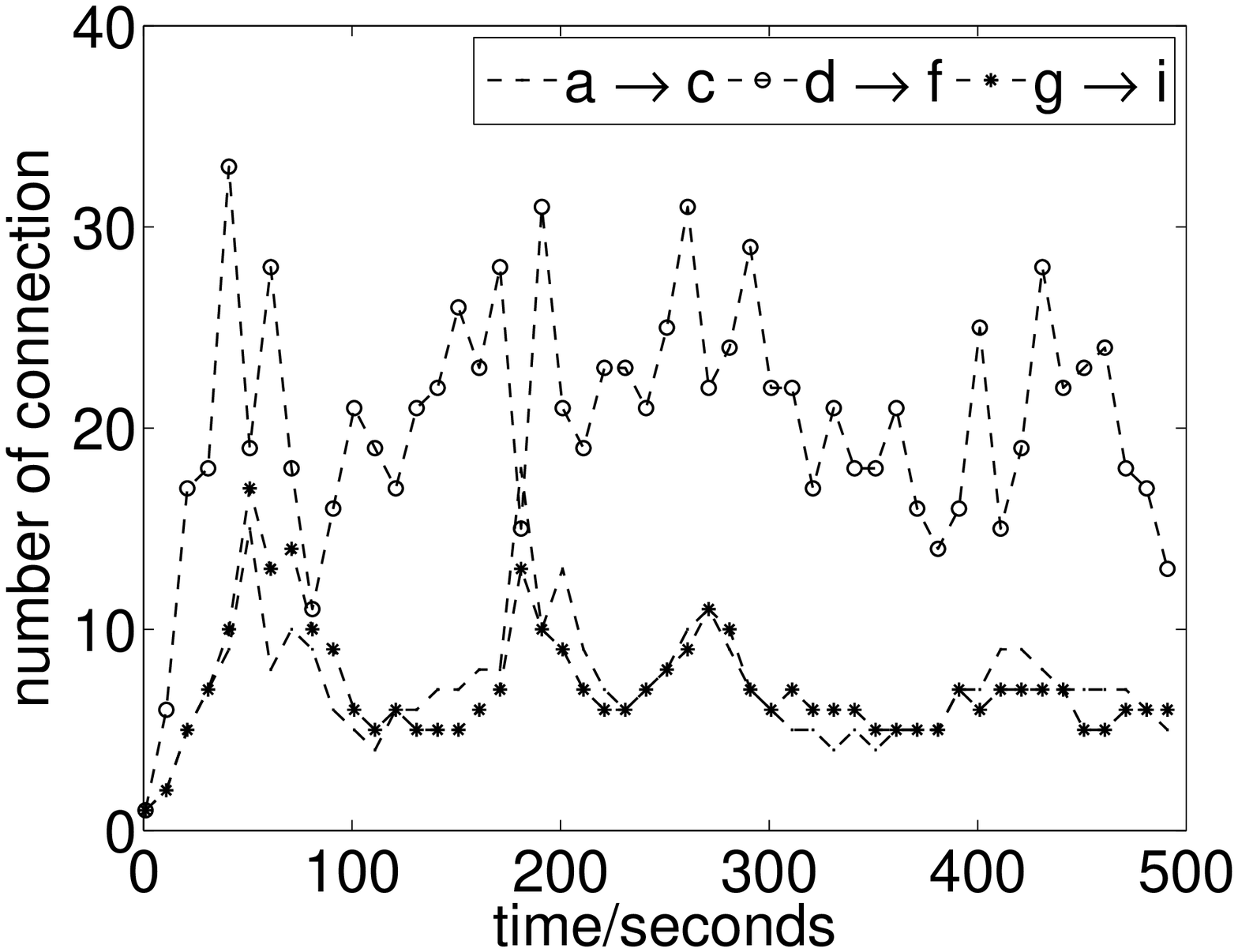}
   \label{fig:nc.mh}
   }
   \caption{Multihop networks scenarios: (a) topology $d$ and its associated conflict graph. (b) throughput (c) curves of transmission aggressiveness. (d) curves of number of connections.}
\end{figure*}

In this part, we consider multiple hop networks scenario. The simulated topology and its associated conflict graph under study is shown in Fig.~\ref{fig:to.mh}. It is a nine-node wireless network with six links represented by dashed line. There are three user flows whose paths are represented by solid lines.

Fig.~\ref{fig:thr.mh} plots the optimal throughput and throughput achieved by multi-connection TCP Reno scheme and TCP Reno over L-CSMA. The middle TCP flow starves when running TCP Reno over L-CSMA. In contrast, multi-connection TCP Reno over A-CSMA with AQM scheme assigns $83$ packets to middle TCP flow, which is within $25\%$ of the optimal achievable rate for this topology.

Fig.~\ref{fig:ta.mh} and \ref{fig:nc.mh} give the evolution of transmission aggressiveness $r$ and number of connection $n_s$. They are both stable in this scenario. The $r$ is mostly around $0.003$, therefore, the dropping probability is not large.

The results indicate that our multi-connection TCP over A-CSMA with AQM scheme is also applicable for multiple hop networks scenario.

\subsection{Multihop Networks with Wireless and Wired Links Scenario}
\begin{figure}[htbp]
   \centering
\subfigure[]{
\includegraphics[scale=0.6]{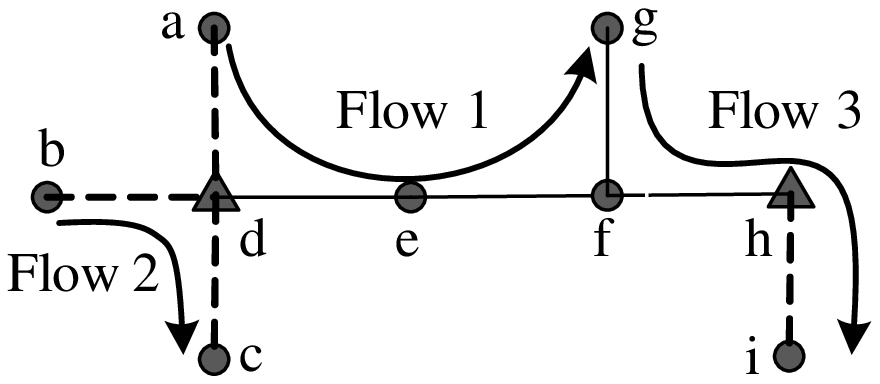}
\label{fig:to.hy}
}
\subfigure[]{
   \includegraphics[scale=0.28]{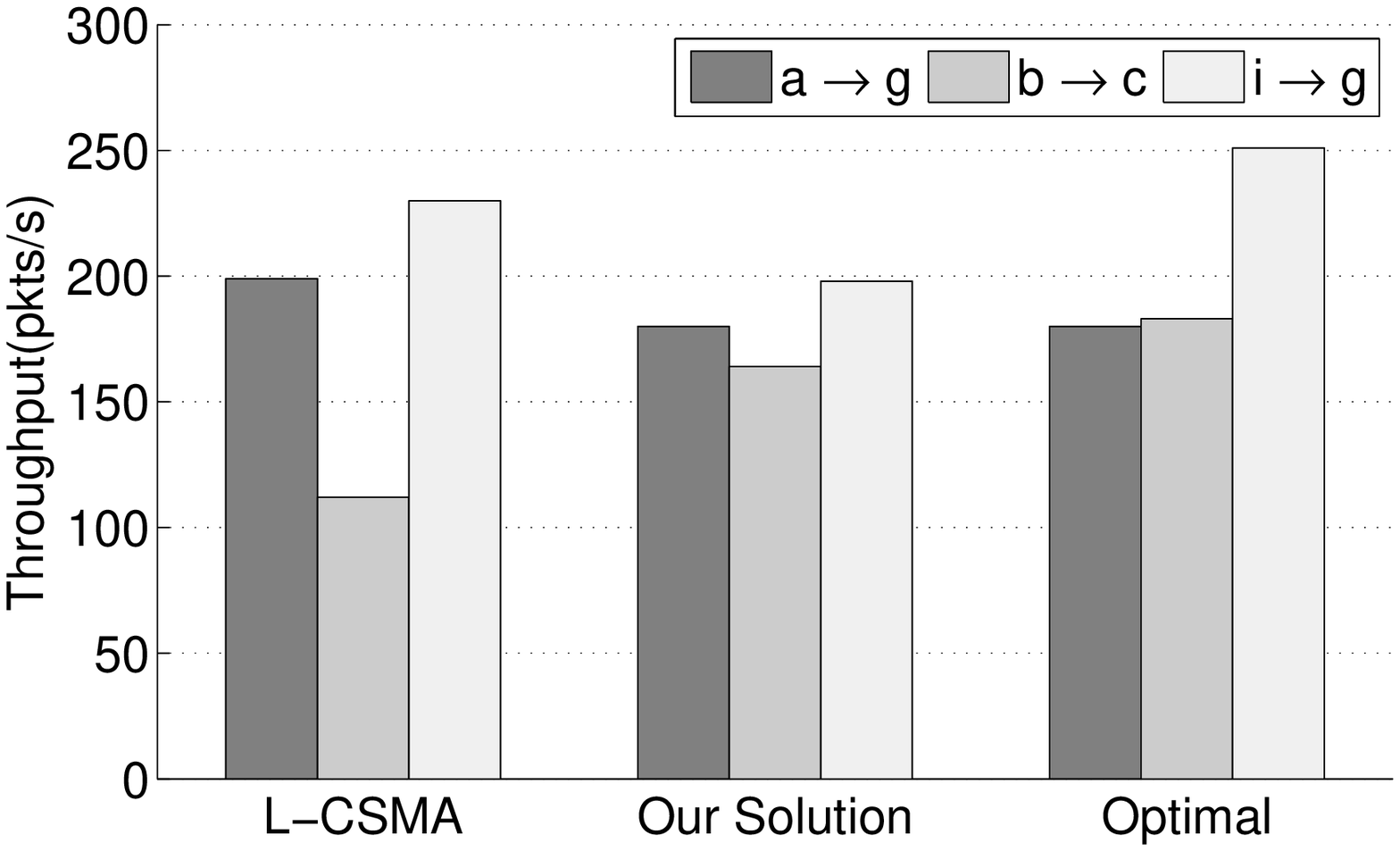}
   \label{fig:thr.hy}
}
\caption{Multihop networks with wired and wireless links scenario: (a) topology $e$ (b) throughput}
\end{figure}

In this subsection, we validate our multi-connection TCP Reno solution for multihop networks with wired and wireless links scenario. The simulated topology under study is shown in Fig.~\ref{fig:to.hy}. It is a nine-node wireless network with two wireless APes denoted by node $d$ and node $h$. Wireless client nodes $a$, $b$ and $c$ are associated with AP $d$ while wireless client node $i$ is associated with AP $h$. Nodes $e$, $f$ and $g$ are wired nodes. In the graph, wireless links are represented by dashed line and wired links are represented by solid line. There are three user flows whose paths are from node $a$ to node $g$, node $b$ to node $c$ and node $i$ to $g$ respectively. The wired link $f$-to-$g$ has the link capacity of $2$Mb and other wired links have the link capacity of $20$ Mb. So the wired bottleneck is on the link $f$-to-$g$. The wireless link bandwidth is $11$ Mb. Other simulation parameters are the same as the previous simulations.

Fig.~\ref{fig:thr.hy} plots the simulation throughput and optimal throughput. We observe TCP Reno does not suffer server starvation over L-CSMA in this topology. However, the utility gap of our solution is $-0.9$ which is closer to zero than that of TCP Reno over L-CSMA, whose utility gap is $-1.8$. This result validates our solution has better performance than that of TCP Reno over L-CSMA. The results validate multi-connection TCP Reno can work properly in multihop networks with wired and wireless links.


\section{Conclusion}

\label{sec:conclu} In this paper, we study whether the widely-installed
TCP Reno is compatible with adaptive CSMA. We find that running TCP
Reno directly over adaptive CSMA suffers from the same severe starvation
problems as TCP Reno over legacy CSMA (IEEE 802.11). The reason
the potentials of adaptive CSMA cannot be realized by TCP Reno is
that the overall system is not stable.

We then propose a multi-connection TCP Reno solution to inter-work with
A-CSMA in a compatible manner. By adjusting the number of TCP connections for each session, our solution can achieve arbitrary system utility. 

Our solution can be implemented at the application layer or transport
layer. Application layer implementation requires no kernel modification,
and the solution can be readily deployed in networks running adaptive
CSMA. Finally, we show that our solution is applicable to single-hop
wireless networks as well as multihop networks with wired and wireless
links.


\bibliographystyle{IEEEtran}

\begin{thebibliography}{10}
\providecommand{\url}[1]{#1}
\csname url@samestyle\endcsname
\providecommand{\newblock}{\relax}
\providecommand{\bibinfo}[2]{#2}
\providecommand{\BIBentrySTDinterwordspacing}{\spaceskip=0pt\relax}
\providecommand{\BIBentryALTinterwordstretchfactor}{4}
\providecommand{\BIBentryALTinterwordspacing}{\spaceskip=\fontdimen2\font plus
\BIBentryALTinterwordstretchfactor\fontdimen3\font minus
  \fontdimen4\font\relax}
\providecommand{\BIBforeignlanguage}[2]{{%
\expandafter\ifx\csname l@#1\endcsname\relax
\typeout{** WARNING: IEEEtran.bst: No hyphenation pattern has been}%
\typeout{** loaded for the language `#1'. Using the pattern for}%
\typeout{** the default language instead.}%
\else
\language=\csname l@#1\endcsname
\fi
#2}}
\providecommand{\BIBdecl}{\relax}
\BIBdecl

\bibitem{wang2005throughput}
X.~Wang and K.~Kar, ``{Throughput modelling and fairness issues in CSMA/CA
  based ad-hoc networks},'' in \emph{Proceedins of IEEE INFOCOM}, 2005.

\bibitem{liew1854back}
\BIBentryALTinterwordspacing
S.~Liew, C.~Kai, J.~Leung, and B.~Wong, ``{Back-of-the-envelope computation of
  throughput distributions in CSMA wireless networks},'' \emph{IEEE Trans
  Mobile Computing}, 2010. [Online]. Available:
  http://arxiv.org//pdf/0712.1854
\BIBentrySTDinterwordspacing

\bibitem{jiang2008distributed}
L.~Jiang and J.~Walrand, ``{A distributed CSMA algorithm for throughput and
  utility maximization in wireless networks},'' in \emph{Proceedings of
  Communication, Control, and Computing}, 2008, pp. 1511--1519.

\bibitem{liu2010towards}
J.~Liu, Y.~Yi, A.~Proutiere, M.~Chiang, and H.~Poor, ``{Towards utility-optimal
  random access without message passing},'' \emph{Wireless Communications and
  Mobile Computing}, pp. 115--128, 2010.

\bibitem{ni2010q}
J.~Ni, B.~Tan, and R.~Srikant, ``{Q-CSMA: Queue-Length based CSMA/CA algorithms
  for achieving maximum throughput and low delay in wireless networks},'' in
  \emph{IEEE INFOCOM Mini-Conference}, 2010.

\bibitem{rajagopalan2008distributed}
S.~Rajagopalan and D.~Shah, ``{Distributed algorithm and reversible network},''
  in \emph{Proceedings of CISS}, 2008, pp. 498--502.

\bibitem{chen2010markov}
M.~Chen, S.~Liew, Z.~Shao, and C.~Kai, ``{Markov approximation for
  combinatorial network optimization},'' in \emph{Proceedings of IEEE INFOCOM},
  2010.

\bibitem{lee2009implementing}
J.~Lee, J.~Lee, Y.~Yi, S.~Chong, A.~Proutiere, and M.~Chiang, ``{Implementing
  Utility-Optimal CSMA},'' in \emph{Proceedings of Allerton Conference}.\hskip
  1em plus 0.5em minus 0.4em\relax Citeseer, 2009.

\bibitem{rangwala2009neighborhood}
S.~Rangwala, A.~Jindal, K.~Jang, K.~Psounis, and R.~Govindan,
  ``{Neighborhood-centric congestion control for multi-hop wireless mesh
  networks},'' \emph{IEEE/ACM Transactions on Networking}, 2009.

\bibitem{tan2007congestion}
K.~Tan, F.~Jiang, Q.~Zhang, and X.~Shen, ``{Congestion control in multihop
  wireless networks},'' \emph{IEEE Transactions on Vehicular Technology}, pp.
  863--873, 2007.

\bibitem{chen2006flow}
M.~Chen and A.~Zakhor, ``{Flow control over wireless network and application
  layer implementation},'' in \emph{Proceedings of IEEE INFOCOM}, 2006.

\bibitem{tullimas2008multimedia}
S.~Tullimas, T.~Nguyen, R.~Edgecomb, and S.~Cheung, ``{Multimedia streaming
  using multiple TCP connections},'' \emph{ACM TOMCCAP}, pp. 1--20, 2008.

\bibitem{lin2004joint}
X.~Lin and N.~Shroff, ``{Joint rate control and scheduling in multihop wireless
  networks},'' in \emph{IEEE CDC}, 2004.

\bibitem{chen2005joint}
L.~Chen, S.~Low, and J.~Doyle, ``{Joint congestion control and media access
  control design for ad hoc wireless networks},'' in \emph{Proceedings of IEEE
  INFOCOM}, 2005.

\bibitem{chen2006multiple}
M.~Chen and A.~Zakhor, ``{Multiple TFRC connections based rate control for
  wireless networks},'' \emph{IEEE Transactions on Multimedia}, pp. 1045--1062,
  2006.

\bibitem{floyd1993random}
S.~Floyd and V.~Jacobson, ``{Random early detection gateways for congestion
  control},'' \emph{IEEE/ACM Transactions on Networking}, pp. 397--412, 1993.

\bibitem{clark1998explicit}
D.~Clark and W.~Fang, ``{Explicit allocation of best-effort packet delivery
  service},'' \emph{IEEE/ACM Transactions on Networking}, pp. 362--373, 1998.

\bibitem{jiang2008improving}
L.~Jiang and S.~Liew, ``{Improving throughput and fairness by reducing exposed
  and hidden nodes in 802.11 networks},'' \emph{IEEE Transactions on Mobile
  Computing}, pp. 34--49, 2008.

\bibitem{chau2009capacity}
C.~Chau, M.~Chen, and S.~Liew, ``{Capacity of large-scale CSMA wireless
  networks},'' in \emph{Proceedings of MobiCom}, 2009, pp. 97--108.

\bibitem{jain2005impact}
K.~Jain, J.~Padhye, V.~Padmanabhan, and L.~Qiu, ``{Impact of interference on
  multi-hop wireless network performance},'' \emph{Wireless Networks}, pp.
  471--487, 2005.

\bibitem{baker1994npc}
B.~Baker, ``Approximation algorithms for np-complete problems on planar
  graphs,'' \emph{Journal of the ACM}, pp. 153--180, 1994.

\bibitem{boorstyn1987throughput}
R.~Boorstyn, A.~Kershenbaum, B.~Maglaris, and V.~Sahin, ``{Throughput analysis
  in multihop CSMA packet radio networks},'' \emph{IEEE Transactions on
  Communications}, pp. 267--274, 1987.

\bibitem{network.optimization.control}
S.~Shakkottai and R.~Srikant, ``Network optimization and control,'' in
  \emph{Foundations and Trends in Networking}, 2007, pp. 271--379.

\bibitem{tang2007equilibrium}
A.~Tang, J.~Wang, S.~Low, and M.~Chiang, ``{Equilibrium of heterogeneous
  congestion control: Existence and uniqueness},'' \emph{IEEE/ACM Transactions
  on Networking}, p. 837, 2007.

\bibitem{mo2000fair}
J.~Mo and J.~Walrand, ``{Fair end-to-end window-based congestion control},''
  \emph{IEEE/ACM Transactions on Networking}, pp. 556--567, 2000.

\bibitem{jiang2009convergence}
L.~Jiang and J.~Walrand, ``{Convergence and stability of a distributed CSMA
  algorithm for maximal network throughput},'' \emph{UC Berkeley, Mar}, pp.
  2009--43, 2009.

\bibitem{shao2010cross}
Z.~Shao, M.~Chen, A.~S. Avestimehr, and S.~Li, ``{Cross-layer Optimization for
  Wireless Networks with Deterministic Channel Models},'' in \emph{Proceedings
  of IEEE INFOCOM}, 2010.

\end{thebibliography}

\appendix

\subsection{Starvation of TCP Reno over L-CSMA}
\label{sec:tcp.lcsma}
We will explain why starvation happens in L-CSMA (CSMA/CA) networks. In L-CSMA (IEEE 802.11), all link has equal transmission aggressiveness $r_l$. Hence, for simplicity, we define $\rho=\exp(\beta r_l)$. From \eqref{eq:csma.stable.rate},
the link throughput achieved by L-CSMA is then given by \begin{equation}
y_{l}=\frac{\sum_{i:l\in i}\left(\prod_{l\in i}\rho\right)}{\sum_{i^{\prime}\in\mathcal{I}}\left(\prod_{l\in i^{\prime}}\rho\right)},\;\forall l\in E.\label{eq:lcsma.rate}\end{equation}
As seen from \eqref{eq:lcsma.rate}, L-CSMA only can schedule a
small fraction of capacity region, and is therefore not throughput-optimal.

We use a simple example to illustrate the starvation of TCP over CSMA/CS networks and explain why this problem remains unsolved. Fairness is one of the key problems that must be considered in designing any MAC, which allows contending links to share the wireless channel fairly. The random backoff algorithm in the L-CSMA network gives each host equal average count-down counter to grab the channel during transmission. This random count-down algorithm works fine in a symmetric environment where all hosts connected to a single access point. However, in asymmetric settings, it fails to achieve the fair channel accessing. Even though a fair higher layer rate control protocol such as TCP cannot solve this MAC layer fairness problem.

Consider the example shown in Fig.~\ref{fig:simple.case}, where each of the four links
runs L-CSMA MAC algorithm. The left hand side figure shows four nearby WLANs, whose access points denoted by triangle each connects to a single wireless client denoted by a circle. The right hand side figure gives the conflict graph of the networks. There is one TCP Reno session running over each link. This network topology is first studied in ~\cite{liew1854back}. We run NS2 simulations to study the rate of each TCP Reno flows. The data rate of L-CSMA is $11$Mbps and packet payload is $1000$Byts in simulation. Simulation results show that rate of link $2$ is nearly zero, while rates of other links get higher rates.

%

This starvation is caused by the fundamental inefficiency of L-CSMA,
and can be intuitively explained as follows. When link $1$ is transmitting,
link $2$ will freeze its count down process because it is within
the carrier sensing range of link $1$. While link $2$ is frozen,
either link $3$ or link $4$ can transmit. This will continuously
freeze link $2$ after link $1$ finishes its transmission. Link $1$,
link $3$ and $4$ take turns to occupy the channel, leaving link
$2$ very small chance to transmit.

In particular, the independent sets for
the example in Fig.\ref{fig:simple.case} are $\emptyset,\{1\},\{2\},\{3\},\{4\},\{1,3\},\{1,4\}$,
and link $2$'s throughput can be computed by \eqref{eq:lcsma.rate}
as follows \begin{equation}
y_{2}=\frac{\rho}{2\rho^{2}+4\rho+1}.\label{eq:link2.rate}\end{equation}
where $\rho\approx2.24$ for networks running L-CSMA under practical
setting\cite{wang2005throughput}. Plugging $\rho$ into \eqref{eq:link2.rate},
we observe that $y_{2}\approx0.11$, which is much smaller compared
to the throughput of other links. As such, the TCP Reno
session running over link $2$ starves. For more discussions
on the starvation of TCP Reno over L-CSMA, please refer to\cite{wang2005throughput}~\cite{liew1854back}\cite{rangwala2009neighborhood}
\cite{tan2007congestion}.

\subsection{TCP Reno over A-CSMA with AQM}
\label{sec:acsma.aqm} In this solution, the only modification needed is each transmitter
of link $l$ applies an AQM policy that each link drops the incoming packet with probability proportional
to $r_{l}$.

\subsubsection{TCP Reno starves}
\begin{figure}[htb]\centering
\centering
  \subfigure[Throughput]{\label{fig:tcp.acsma.throughput}\includegraphics[scale=0.3]{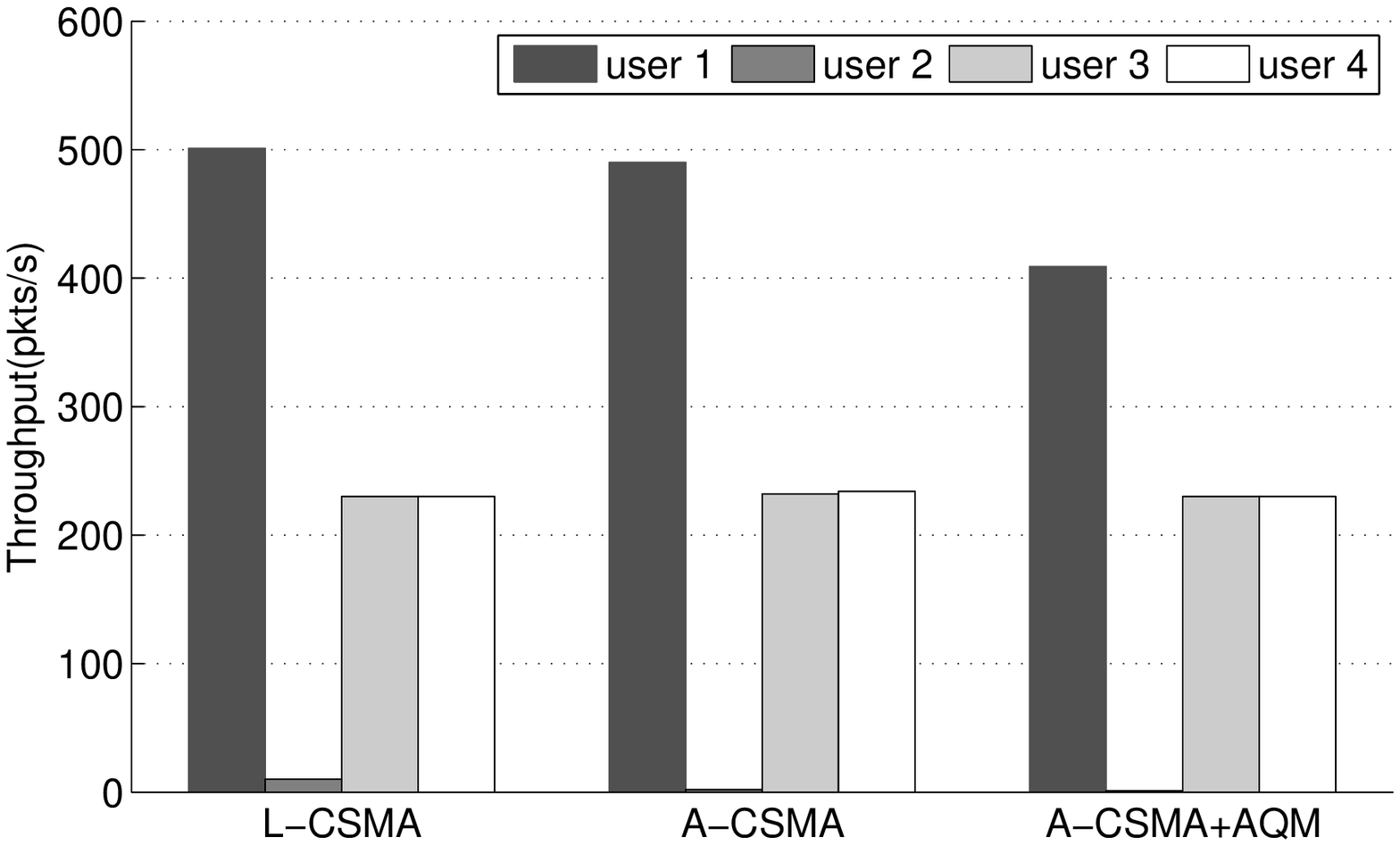}}
  \subfigure[Transmission Aggressiveness]{\label{fig:tcp.acsma.aqm.r}\includegraphics[scale=0.22]{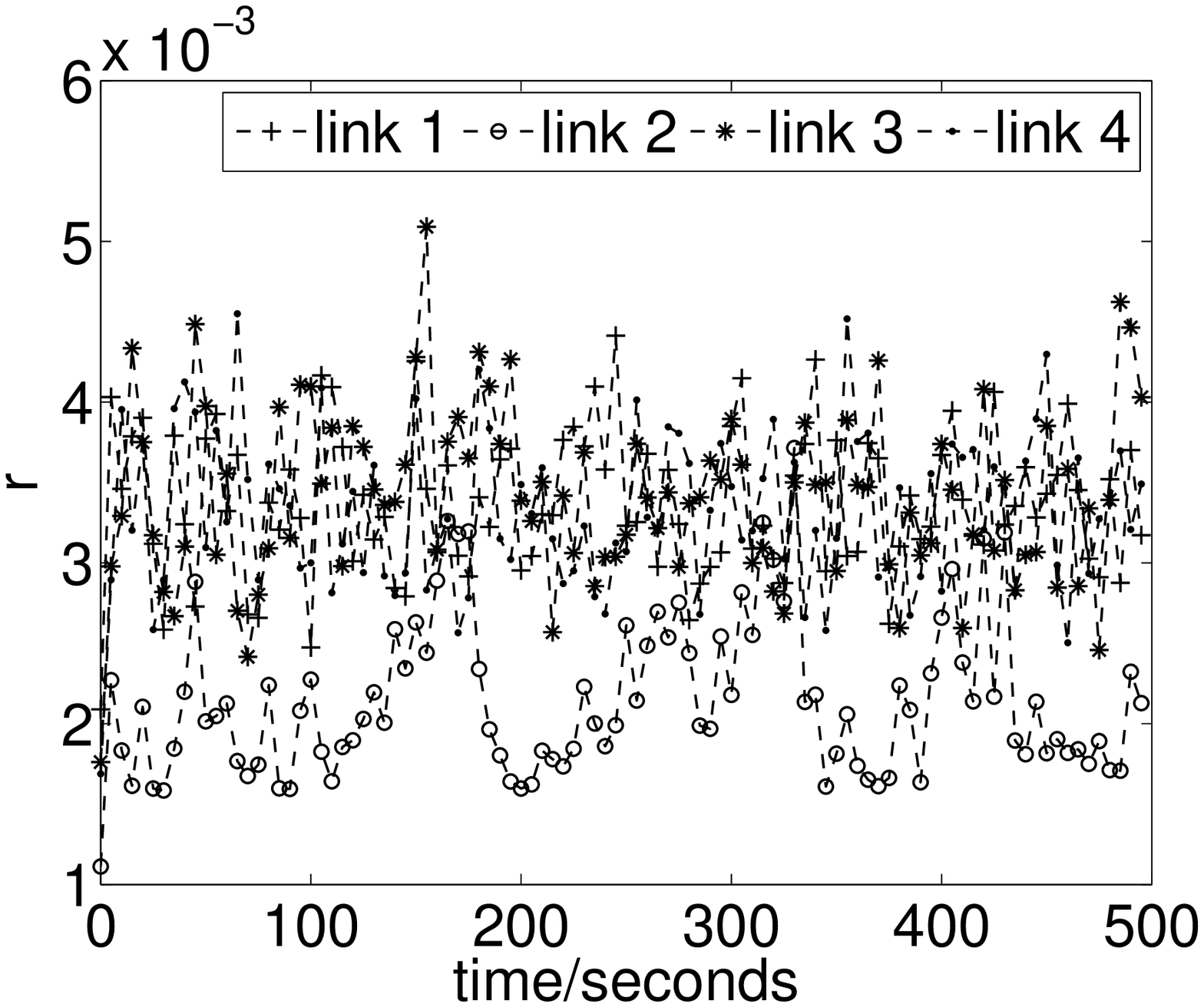}}
  \subfigure[Round-trip-time]{\label{fig:tcp.acsma.aqm.rtt}\includegraphics[scale=0.22]{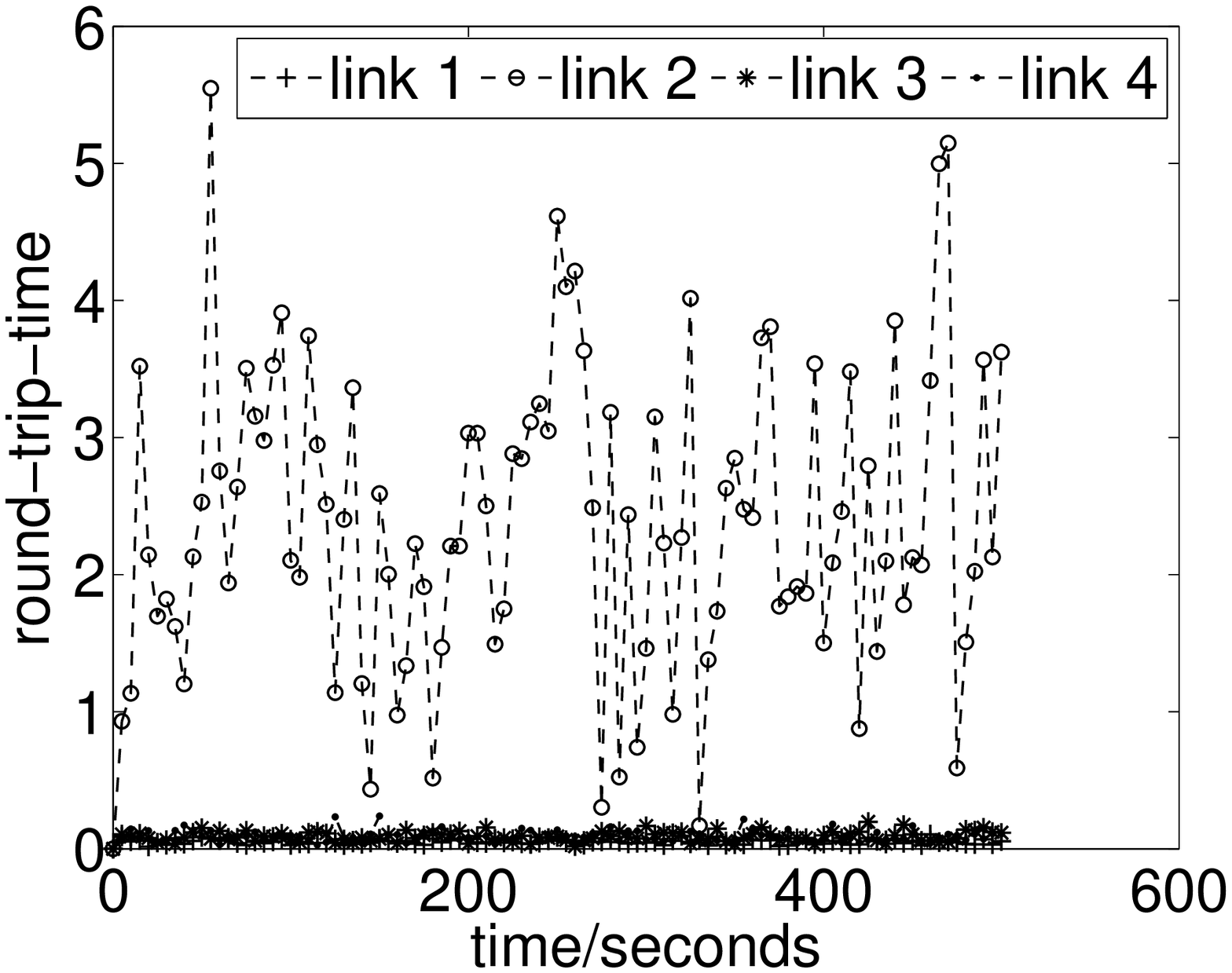}}
  \caption{Simulation Results: TCP Reno over A-CSMA with AQM.}
  \label{fig:result.tcp.acsma.aqm}
\end{figure}

We conduct
an NS-2 simulation to demonstrate the problem. The simulation topology
and setup are the same as that in Fig. \ref{fig:simple.case}. But
each link $l$ drops incoming packets with probability proportional
to $r_{l}$. We also give the backward links higher priority.
The simulation results are shown in Fig. \ref{fig:tcp.acsma.throughput}, from which we can see that user $2$ starves. Fig. \ref{fig:tcp.acsma.aqm.r} and \ref{fig:tcp.acsma.aqm.rtt}
show the transmission aggressiveness of the links, and RTT of TCP Reno
sessions, respectively. We observe that all links have closely equal transmission aggressiveness. Under this situation, by now it should be clear that link $2$ will starve.

\subsubsection{Explanation}
We present the analysis as follows.
In this example, there are four TCP Reno sessions $s=1,2,3,4$, each running over a wireless link $l=1,2,3,4$.
With TCP Reno as the rate control algorithm, the algorithms in (\ref{eq:mtcp})-(\ref{eq:r}) turn into:
\begin{numcases}{}
   \dot{x_s}= \frac{x_{s}^{2}}{2}\left(\frac{2}{T_{s}^{2}x_{s}^{2}}-r_{s}\right)_{x_s}^+, \;\forall s=1,2,3,4; \label{eq:tcp1} \\
   \dot{r_s}= \alpha \big[x_s - \sum_{i:s\in i}\tau_i (\br) \big]_{r_s}^+, \; \forall s=1,2,3,4; \label{eq:acsma1} \\
   \tau_i (\br) = \frac{\exp\left( \sum_{s\in i} \beta r_s\right)}{\sum_{i^{\prime}\in \mathcal{I}}\exp\left( \sum_{s\in i^{\prime}} \beta r_s\right)}, \; \forall i\in \mathcal{I}; \label{eq:br1}\\
   T_s =  \frac{1}{\alpha}\frac{r_s}{x_s}, \; \forall s=1,2,3,4. \label{eq:rtt}
\end{numcases}
The first equation is the TCP Reno updating equation, as discussed in (\ref{eq:tcp.rate}). The second and the third equations correspond to A-CSMA scheduling and the transmission aggressiveness adjusting equation.

The fourth equation characterizes the RTT user $s$ experiences. As we mentioned in the simulation settings, to simplify the analysis, we only consider the forward link by giving the TCP ACK higher priority to transmit. Therefore, the waiting time of TCP ACK is very short. Most of the packets are buffered at the queue of forward link. Thus, $W_s \approx Q_s$ (recall $W_s$ is the congestion window size of TCP Reno session of user $s$), and round trip time $T_s$ can be written as
\begin{equation}
   T_s = \frac{W_s}{x_s} \approx \frac{Q_s}{x_s},
\end{equation}
where $Q_s$ is the queue length of link $s$ used by user $s$. According to A-CSMA, the transmission aggressiveness is proportional to the queue length. Thus we obtain (\ref{eq:rtt}):
\begin{equation}\label{eq:rlql}
    T_s =  \frac{Q_s}{x_s} = \frac{1}{\alpha}\frac{r_s}{x_s}, \; \forall s=1,2,3,4.
\end{equation}

Note the algorithms in (\ref{eq:tcp1})-(\ref{eq:rtt}) are a special case of the generic one in (\ref{eq:mtcp})-(\ref{eq:r}), except the last RTT equation in (\ref{eq:rtt}). The RTT equation reveals a subtle interaction between TCP Reno and A-CSMA, which in fact causes the starvation. We explain it as below.

At the equilibrium of the dynamic system described in (\ref{eq:tcp1})-(\ref{eq:rtt}), the derivatives are all zero. Hence,
\begin{eqnarray}\label{eq:xsrl}
    r_s &=& \frac{2}{T_s^2 x_s^2}, \; \forall s=1,2,3,4.
\end{eqnarray}

Solving (\ref{eq:rlql}) and (\ref{eq:xsrl}), we arrive at the following observations: \begin{eqnarray}
r_s &=& 2^{1/3}\alpha^{2/3}, \; \forall s=1,2,3,4.
\label{eq:convg_r}
\end{eqnarray}
Thus every link $s$ has the same $r_s$ and competes the channel
with the same level of transmission aggressiveness. Under this
situation, our previous discussions indicate that links 1, 3 and 4 will take turns to freeze link 2, and hence the TCP Reno over link 2 will starve.

\subsection{Proof to Theorem \ref{the:sol}}
\label{sec:proof}
We give the proof of Theorem \ref{the:sol} as follows:

Proof:
1) By relaxing the first set of inequalities (\ref{eq:flow}) of problem \textbf{MP}, we get its partial Lagrangian as follows:
\begin{equation}
\begin{array}{rl}
   L(\bx, \btau, \br) =& \sum_{s\in \mathcal{S}} U_s(x_s) -\frac{1}{\beta} \sum_{i\in\mathcal{I}}\tau_{i}\log\tau_{i} \\
   &- \sum_{l\in E}r_l \left(\sum_{s:l\in s, s\in \mathcal{S}}x_s - \sum_{i:l \in i}\tau_i \right),
\end{array}
\end{equation}
where $\br = [r_l, l \in E]$ is the vector of Lagrange multipliers. We notice that
$\sum_{l\in E} r_l\sum_{i:l \in i}\tau_i = \sum_{i \in \mathcal{I}}\tau_i \sum_{l\in i} r_l$.
Since the problem \textbf{MP} is a concave optimization problem and Slater's condition holds, the optimal solution of problem \textbf{MP} can be found by solving the following problem successively in $\btau$, $\bx$ and $\br$:
\begin{equation}
\begin{array}{rl}\label{pro:lag}
   \displaystyle\min_{\br \geq 0} \max_{\bx \geq 0,\btau \geq 0}
   &\displaystyle\sum_{s\in \mathcal{S}} U_s(x_s) - \sum_{l\in E} r_l \sum_{s:l\in s}x_s\\
   &+ \sum_{i\in \mathcal{I}}\tau_i \sum_{l\in i} r_l - \frac{1}{\beta} \sum_{i\in\mathcal{I}}\tau_{i}\log\tau_{i} \\
   \mbox{s.t.}
   & \displaystyle\sum_{i\in \mathcal{I}} \tau_i = 1.
   \end{array}
\end{equation}
Define
\begin{equation}
\label{eq:1}
   g_{\beta}(\br) \triangleq \frac{1}{\beta}\log\left(\sum_{i \in \mathcal{I}} \exp(\beta \sum_{l\in i}r_l)\right).
\end{equation}
The conjugate of $g_{\beta}(\by)$ is defined as $g^*(z) = \sup_{\by\in dom g} (z^T-g(\by))$. Therefore, the conjugate of $g_{\beta}(\br)$ is given by
\begin{numcases}{g^*_{\beta}(\btau)=}
    \frac{1}{\beta}\sum_{i\in \mathcal{I}} \tau_i\log \tau_i & if $\btau \geq 0$ and $1^T \btau = 1$\\
    \infty & otherwise.
\end{numcases}
The conjugate of its conjugate is itself, i.e., $g_{\beta}(\br)= g^{**}_{\beta}(\br)$. Therefore, we have
\begin{equation}
\label{eq:2}
\begin{array}{rl}
g_{\beta}(\br) = \max_{\btau \geq 0}
    & \displaystyle\sum_{i\in \mathcal{I}}\tau_i\sum_{l\in i}r_l - \frac{1}{\beta}\sum_{i\in \mathcal{I}} \tau_i\log \tau_i \\
    \mbox{s.t.} & \displaystyle\sum_{i\in \mathcal{I}} \tau_i = 1,
\end{array}
\end{equation}
with the corresponding unique optimal solution
\begin{equation}
   \tau_i(\br)=\frac{\exp\left(\sum_{l\in i}\beta r_l\right)}{\sum_{i^{\prime}\in\mathcal{I}}\exp\left(\sum_{l\in i^{\prime}}\beta r_l\right)},\;\forall i\in \mathcal{I}. \label{eq:sta}
\end{equation}
(Note this is the exact stationary distribution that CSMA Markov chain achieves. Hence, A-CSMA can be used to solve this subproblem.)
From (\ref{eq:1}) and (\ref{eq:2}), formula (\ref{pro:lag}) will be
\begin{eqnarray}\label{pro:lag.approx}
   \min_{\br\geq 0} \max_{\bx \geq 0} \sum_{s\in \mathcal{S}} U_s(x_s)- \sum_{l\in E}r_l \sum_{s:l\in s}x_s  \nonumber \\
   + \frac{1}{\beta}\log\left(\sum_{i \in \mathcal{I}} \exp(\beta \sum_{l\in i}r_l)\right).
\end{eqnarray}
Define
\begin{eqnarray}
   L_2(\bx,\br) \triangleq &\sum_{s\in \mathcal{S}} U_s(x_s)
    - \sum_{l\in E}r_l \sum_{s:l\in s}x_s  \nonumber s\\
    &+ \frac{1}{\beta}\log\left(\sum_{i \in \mathcal{I}} \exp(\beta \sum_{l\in i}r_l)\right).
\end{eqnarray}
The saddle point of above formula is the optimal primal and dual variable. Therefore, the partial derivative is equal to zero, we have
\begin{eqnarray}
   \frac{\partial L_2(\bx,\br)}{\partial x_s} &=& U_s^{\prime}(x_s)- \sum_{l \in s} r_l = 0,\\
   \frac{\partial L_2(\bx,\br)}{\partial r_l} &=& -r_l\left( \sum_{s:l\in s}x_s - \sum_{i:l\in i}\tau_i (\br)\right)=0.
\end{eqnarray}
When $U_s(x_s) = \frac{1}{x_s}$, a primal-dual algorithm that solves above problem is
\begin{numcases}{}
   \dot{x_s}= f(x_s)\left(\frac{2n^2_s}{T_{s}^{2}x_{s}^{2}}-\sum_{l\in s}r_{l}\right)_{x_s}^+,\;\forall s\in \mathcal{S};\\
   \dot{r_l}= \alpha(r_l) \big[\sum_{s:l\in s}x_s - \sum_{i:l\in i}\tau_i (\br) \big]_{r_l}^+, \;\forall l\in L.
\end{numcases}
Plus the (\ref{eq:sta}), we now have that the equilibrium of dynamic system in (\ref{eq:ns})-(\ref{eq:r}) solves \textbf{MP}.

2) The proof can use the same set of standard Lyapunov elaborated in~\cite{chen2010markov}.

\end{document}